%% file: combinedManuscript_noColor.tex
\documentclass[aps,pre,twocolumn,unsortedaddress]{revtex4-1}
\usepackage{amsmath}
\usepackage{amsfonts}
\usepackage{amssymb}
\usepackage{graphicx}
\usepackage[dvipsnames]{xcolor}
\usepackage{float}
\usepackage{siunitx}
\usepackage[version=4]{mhchem}
\usepackage{xr}
\usepackage{physics}

\usepackage{cases}
\usepackage{mathrsfs}
\usepackage{wrapfig}
\usepackage{titlesec}
\usepackage{tocloft}
\usepackage{mwe}
\usepackage{booktabs}
\usepackage[utf8]{inputenc}
\usepackage{units}
\usepackage{braket}
\usepackage{float}
\usepackage{placeins}
\usepackage[overload]{empheq}

\DeclareSIUnit[number-unit-product = {\,}] \cal{cal}

\newcommand{\Deff}{D_{\text{eff}}}
\newcommand{\kB}{k_\mathrm{B}}
\newcommand{\lB}{l_\mathrm{B}}
\newcommand{\lBz}{l_\mathrm{B0}}

\makeatletter
\def\p@subsection{}
\makeatother

\begin{document}

	\title{The physics of cement cohesion}
	
	\author{Abhay Goyal*$^+$}
	\affiliation{Department  of  Physics,  Institute  for Soft  Matter  Synthesis and Metrology, Georgetown  University,  Washington,  D.C. 20057,  USA}
	
	\author{Ivan Palaia*}
	\affiliation{Universit\'e Paris-Saclay, CNRS, LPTMS, 91405, Orsay, France}
	\affiliation{Department of Physics and Astronomy, University College London, London, WC1E 6BT, United Kingdom}
	
	\author{Katerina Ioannidou}
	\affiliation{Laboratoire de M\'ecanique et G\'enie Civil, CNRS, Universit\'e de Montpellier, 34090 Montpellier, France}
	\affiliation{Massachusetts Institute of Technology/CNRS/Aix-Marseille University Joint Laboratory, Cambridge, MA 02139}
	\affiliation{Department of Civil and Environmental Engineering, Massachusetts Institute of Technology, Cambridge, MA 02139}
	
	\author{Franz-Josef Ulm}
	\affiliation{Department of Civil and Environmental Engineering, Massachusetts Institute of Technology, Cambridge, MA 02139}
	
	\author{Henri van Damme}
	\affiliation{{\'E}cole Sup\'erieure de Physique et Chimie Industrielle de la Ville de Paris, 10 rue Vauquelin, 75005 Paris, France}
	
	\author{Roland J.-M. Pellenq}
	\affiliation{Massachusetts Institute of Technology/CNRS/Aix-Marseille University Joint Laboratory, Cambridge, MA 02139}
	\affiliation{Department  of  Physics, Georgetown  University,  Washington,  D.C. 20057,  USA}	
	
	\author{Emmanuel Trizac}
	\affiliation{Universit\'e Paris-Saclay, CNRS, LPTMS, 91405, Orsay, France}
	
	\author{Emanuela Del Gado$^+$}
	\affiliation{Department  of  Physics,  Institute  for Soft  Matter  Synthesis and Metrology, Georgetown  University,  Washington,  D.C. 20057,  USA}
	
	\date{\today}
	\maketitle
	
	{\bf 	
		Cement is one of the most produced materials in the world. A major player in greenhouse gas emissions, it is the main binding agent in concrete, to which it provides a cohesive strength that rapidly increases during setting. Understanding how such cohesion emerges has been a major obstacle to advances in cement science and technology. Here, we combine computational statistical mechanics and theory to demonstrate how cement cohesion results from the organization of interlocked ions and water, progressively confined in nano-slits between charged surfaces of Calcium-Silicate-Hydrates. 
		{ Due to the water/ions interlocking, dielectric screening is drastically reduced and ionic correlations are proven significantly stronger than previously thought, dictating the evolution of the nano-scale interactions during cement hydration}. By developing a quantitative analytical prediction of cement cohesion based on Coulombic forces, we reconcile a novel fundamental understanding of cement hydration with the fully atomistic description of the solid cement paste and open new paths for science and technologies of construction materials.
	}
	
	\section{Introduction}
	Concrete, made by mixing cement with water, sand and rocks, is by far the most used man-made substance on earth. With a world population projected to grow past 9 billion by mid-century, there is need for more and better infrastructure \cite{UNEnvironment2018}, with no other material that can replace concrete to meet our needs for housing, shelter, or bridges. 
However, concrete as it is now is not sustainable, since cement production alone is responsible for significant amounts of man-made greenhouse gases. While even a slight reduction of its carbon footprint will dramatically reduce global anthropogenic CO$_{2}$ emissions, meeting emission-reduction targets for new constructions calls for deeper scientific understanding of cement properties and performance \cite{Habert2020}.
	
	The dissolution of cement grains in water and re-precipitation of various hydration products, with Calcium-Silicate-Hydrates (C--S--H) being the most important \cite{Allen2007, Qomi2014}, drives the setting of cement into a progressively harder solid that binds together concrete and determines its mechanics \cite{Ioannidou2016c}. During this process, strongly cohesive forces develop from the accumulation and confinement of ions in solution between the surfaces of C--S--H, whose surface charge progressively increases over time \cite{Pellenq2004,Plassard2005,Gmira2004}.
	
	Net attractive interactions between equally charged surfaces in ionic solutions are common in colloidal materials or biological systems \cite{Israelachvili2011,Jho2011,Gelbart2000, Levin02,Moreira2001,Samaj2018,Carrier2014}. The comprehensive analytical theory developed nearly a century ago by Derjaguin, Landau, Verwey and Overbeek (DLVO), which relies on a mean-field approximation treating the ions as an uncorrelated continuum, captures some of these cases \cite{Israelachvili2011}. For cement hydration products like C--S--H, however, the DLVO description 
	is inapplicable since the ions in solution are mostly multivalent (such as Ca$^{2+}$) and confined between surfaces {whose surface charge density rapidly reaches values up to $\simeq 3$ to $5 e^-\si{\per \nano \meter^2}$ in the layers of hardened C--S--H \cite{Gmira2004,Qomi2014,Masoumi2019}.} Sure enough, DLVO theory does not predict any cohesion for cement \cite{Gmira2004,Plassard2005}. 
	
	Monte-Carlo simulations of a primitive model for ion confinement (PM), instead, have proven that ions, confined in water between charged surfaces, can induce net attractive forces for a range of surface charge densities relevant to C--S--H \cite{ Pellenq2004, Jonsson2004, Pellenq2008}, due to the correlations that stem from the discrete nature of ions. Nevertheless, these PM studies predict cohesive strengths at most of $\simeq 60$ MPa in clear contradiction with experiments and with the fully atomistic understanding of hardened C--S--H achieved over the last 10 years \cite{Gmira2004,Vandamme2009,Qomi2014,Mishra2017a,Geng2017}. { The cohesive strength of hardened cement is 100 times larger than that, and the presence of water is limited to a few molecules per ion, whereas it is treated as a bulk dielectric continuum in the PM approach. There is therefore a knowledge gap in the fundamental understanding of how nanoscale cohesive forces emerge during cement hydration, as different chemical reactions drive the increase of the surface charge of cement hydrates and the ion confinement \cite{Bullard2011}.}
	
	We now fill this gap with 3D numerical simulations that feature a simple but molecular description of ions and water, providing a quantitative picture of how cement cohesion develops during hydration. When one considers explicitly the role of water, it becomes clear how its capability to restructure and re-orient around the ions drives the optimized organization of interlocked ion-water structures that determine the net cohesive forces and their evolution. 
	
	As ion confinement and surface charge density increase, with {dramatically weakened water dielectric screening}, electrostatic forces and discreteness effects are dramatically amplified and glue together the ion-water-surface assembly into a highly cohesive state. While we quantitatively recover several key experimental findings in real cement \cite{Meral2011,White2015a,Thomas2001,Bordallo2006,Bohris1998,Fratini2013} and the main features of the fully atomistic description of hardened C--S--H \cite{Masoumi2017,Masoumi2019}, we test the emerging physical picture against an analytical theory that distills the essential ingredients of the net interactions beyond the traditional PM assumptions, for a range of materials and systems in similar strong electrostatic coupling conditions.   
	
	\section{Results}
	\begin{figure*}
		\centering
		\includegraphics[width=.9\textwidth]{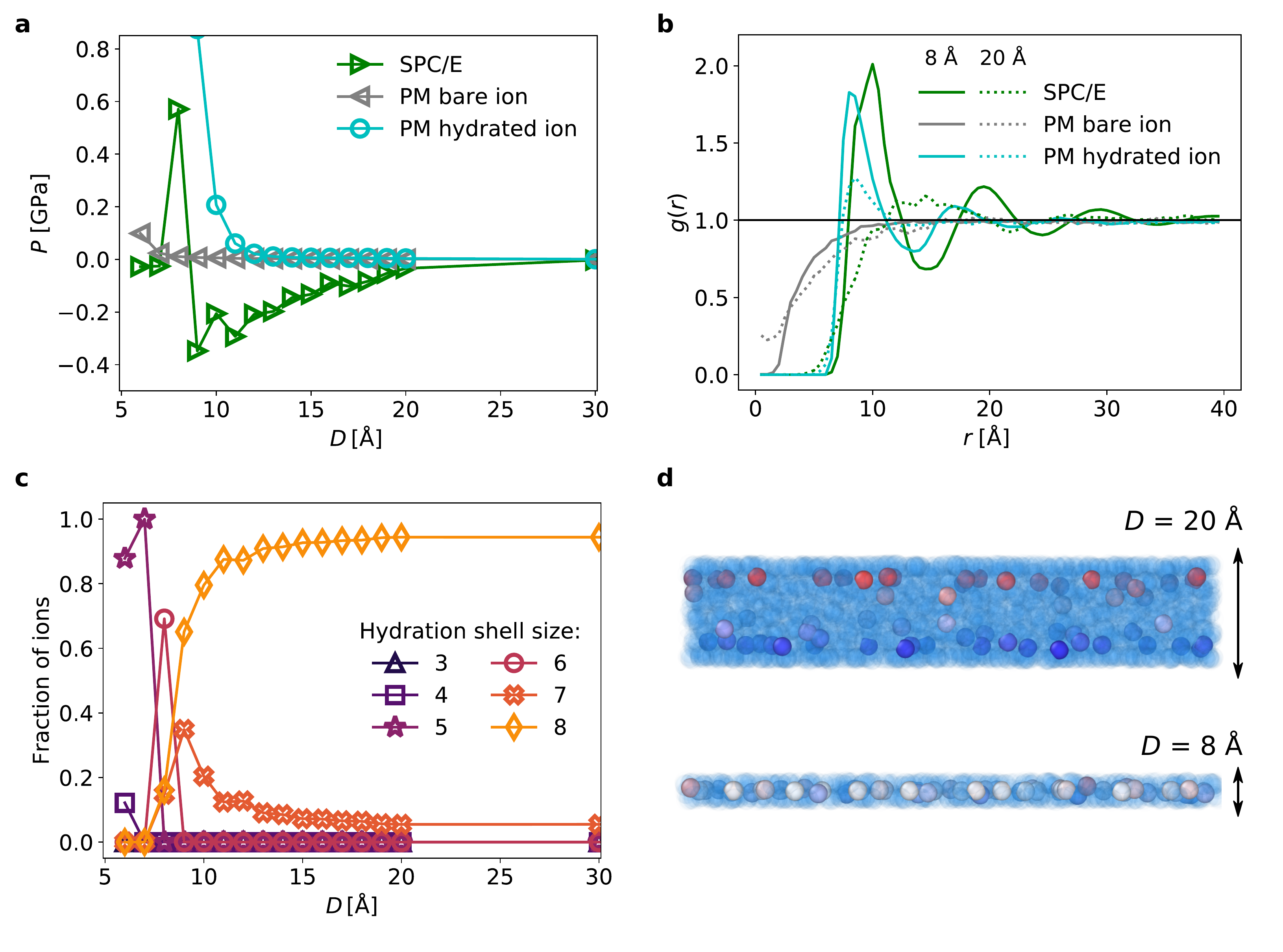}
		\caption{Results from simulations with PM and SPC/E water at $\sigma = 1 e^-\si{\per \nano \meter^2}$. Here and in subsequent figures all data are averaged over $10^{6}$ MD steps after having reached equilibrium. Error bars are smaller than the symbol sizes. In (a) we show the net pressure between the C-S-H surfaces. With explicit water, we obtain negative pressures---i.e. net attraction---that cannot be found with the PM approach, even considering an increased effective ion size due to hydration. This stems from much stronger ion-ion correlations seen in (b), where we plot the ion pair correlation in the $xy$ plane at $D=\rm 8\,\si{\angstrom}$ {and $D=\rm 20\,\si{\angstrom}$}. Using the hydrated ion size in the PM creates a first peak that is similar to what we get with explicit water, but it cannot replicate the long-ranged correlations. The water behavior is more complex than that and strongly depends on confinement. Looking at the hydration shell size (c) at large separation, we observe nearly full hydration shells of 7-8 water molecules. As $D$ is decreased, the shell size does not immediately change and the shells make up a larger portion of the water, which affects its ability to screen electrostatic interactions. When confined to $D=\rm 8\si{\angstrom}$, the ions coalesce into a single layer (d) and the shells are pressed against the walls resulting in a barrier to overcome. }
		\label{lowCharge}
	\end{figure*}
	
	The intricacy of chemical reactions during cement hydration and setting makes it hard to identify the fundamental physical mechanisms that control cement cohesion. C--S--H is a non-stoichiometric compound, with structure and composition variability, even more pronounced at the earlier stages of the hydration \cite{Richardson1993,lothenbach2015,Qomi2014,Geng2017}. The charged surfaces of C--S--H nanoparticles confine ions and water in nanometer sized pores \cite{Pellenq2008,Chiang2012} and studies of titration of the surface silanol groups indicate that during early hydration, as a result of the changing solution chemistry, the surface charge of C--S--H increases with increasing pH, coupled to the combined precipitation of calcium hydroxide \cite{lothenbach2015}. Experimental and simulation efforts over the last decade have clarified the atomistic details of the final hardened C--S--H, in terms of atomic pair distribution functions, composition variability, and even cohesive strength \cite{Meral2011, Qomi2014, White2015a,Geng2017,Mishra2017a}. However, a direct link between the surface charge and chemistry, the emerging nanoscale cohesion, and the final material properties is missing.
	
	To address this question we have used a semi-atomistic computational approach, in which ions and water are represented explicitly while the C--S--H surface properties at different hydration stages are recapitulated through surface charge densities $\sigma$ from $1 e^-\si{\per \nano \meter^2}$ to $3 e^-\si{\per \nano \meter^2}$. { C--S--H is often characterized in terms of Ca/Si ratios, and, with pH values typical of cement hydration, the range of surface charge densities considered here approximately correspond to the relevant range of Ca/Si ratios between 1 and 2 \cite{Richardson1993,Labbez2011,Masoumi2019}. Representing C--S--H surfaces with smooth, uniformly charged walls is clearly a simplification, since their strongly heterogeneous nature is known \cite{Geng2017}, but it is essential to the extended spatio-temporal analysis performed here. Thanks to this simplification, in fact,} we can extensively sample ion and water structure and dynamics in Molecular Dynamics and Grand-Canonical Monte Carlo simulations, and extract the net pressure ions and water induce between the confining C--S--H charged surfaces. 
		
	In our 3D study, C--S--H surfaces are planar (walls), consistent with the platelet-like morphology of the nanoparticles \cite{skinner2010, Chiang2012,Richardson1993} and the ions confined in between them, neutralizing the surface charge, are Calcium (Ca$^{2+}$).To reasonable approximation, this represents the most relevant portion of the ions confined between C--S--H surfaces during cement hydration \cite{Gmira2004,Pellenq2004,Masoumi2019}. The walls have periodic boundaries along $\hat{x}$ and $\hat{y}$ and are separated by a distance $D$ along the $\hat{z}$ direction. For (Ca$^{2+}$) and C--S--H surfaces we include both short-range steric/dispersion and long-range electrostatic forces, as described in the Methods. Explicitly including the molecular degrees of freedom of water {(we use SPC/E and in some cases TIP4P/2005 water, as also described in Methods)} is key to capture its behavior under confinement \cite{Schlaich2016,Giovambattista2009,Fumagalli2018}.
	
	Already for the lowest $\sigma = 1 e^-\si{\per \nano \meter^2}$, corresponding to very early hydration, the inclusion of explicit water leads to a net attraction (Fig. \ref{lowCharge}a) arising from strong and long-ranged correlations in ion positions: the ions are localized in the $z$ direction like in the PM but with explicit water the pair correlation of their positions $g(r)$ in the $xy$ plane has clear peaks that persist to large distances (Fig. \ref{lowCharge}b). The same $g(r)$ for PM (bare ions) indicates instead a spatial arrangement close to uncorrelated, and a simplistic attempt to account for the ions hydration shells by considering a larger effective ion size is still insufficient to capture the long-range effect of water and the related pressure profile.
	
	In the confined space between charged surfaces, the ion hydration shells may differ significantly from those in bulk water. In general, we identify the ion-water structures as $n$-mers, $n$ being the number of water molecules surrounding an ion (see Methods), as in Fig. \ref{lowCharge}c. At relatively large separations of $D>\rm 20\, \si{\angstrom}$, hydration shells have typically 8 water molecules as in bulk water (Fig. \ref{lowCharge}c), consistent with a range of simulations and experiments \cite{Koneshan1998,Megyes2004}. Computing dynamic correlations such as those measured through quasi-elastic neutron scattering (QENS) reveals relaxation times consistent with experimental values \cite{Koneshan1998,Megyes2004}---hydration shells are energetically favored due to the high hydration energy of Ca ions \cite{Smith1977}, but 
	quite dynamical as individual water molecules switch between free and bound states (see SM section \ref{sec:dynamics}). 
	\begin{figure*}
		\centering
		\includegraphics[width=\textwidth]{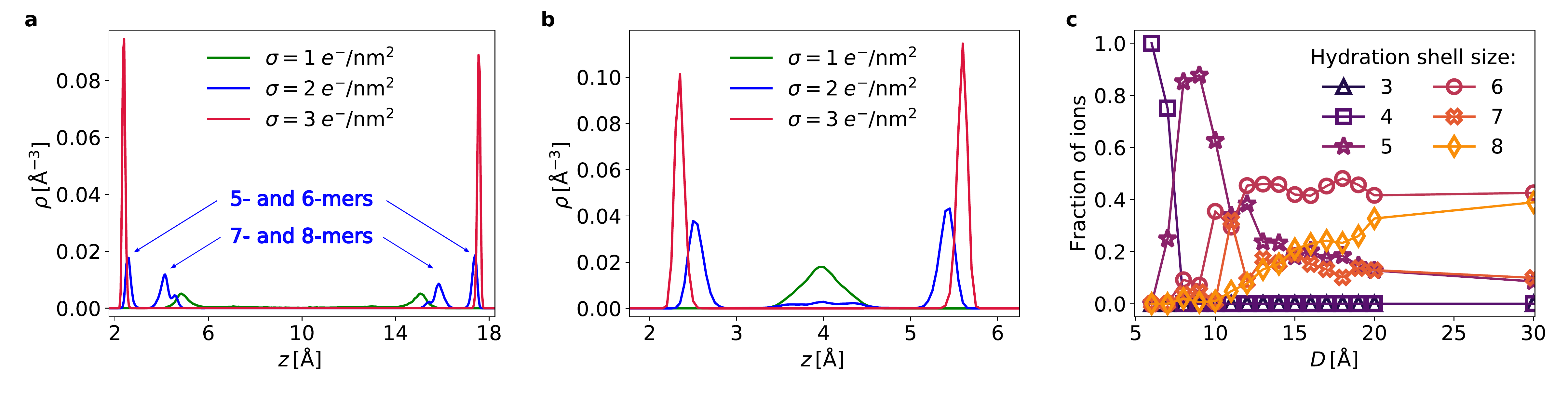}
		\caption{Plots of ion density profiles at (a) $D=\rm 20\,\si{\angstrom}$ and (b) $D=\rm 8\,\si{\angstrom}$. By increasing surface charge density, we see an increasing localization of ions in the $z$ direction. For $\sigma = 2 e^-\si{\per \nano \meter^2}$ at large $D$ (c), there is a split between wall ions with a hemispherical hydration shell of 5-6 water molecules and shifted ions with a nearly full hydration shell of 7-8 water molecules. At lower $D$ or higher $\sigma$, the shifted ions are suppressed and all ions are close to the wall, in stark contrast to the situation at $\sigma = 1 e^-\si{\per \nano \meter^2}$ and $D=\rm 8\,\si{\angstrom}$ where all the ions coalesce into a single layer.}
		\label{ionProfiles}
	\end{figure*}
	As $D$ is decreased, a larger fraction of the total water is in the hydration shells, while ions become increasingly correlated (see the strongly pronounced peaks in the $g(r)$ in Fig. \ref{lowCharge}b). Further confinement, however, starts to dramatically affect the hydration shells---by $D=\rm 8\si{\angstrom}$ the surfaces are squeezing the ions into a single layer and pressing against their hydration shells (Fig. \ref{lowCharge}d), eventually reduced to $\simeq 5$ water molecules per ion. { Hence, reducing $D$ from $9\,\si{\angstrom}$ ({where the full $8$-mer hydration shell can be accommodated}) to $7\,\si{\angstrom}$ (where this is not possible anymore) requires overcoming an energy barrier. }The Calcium ion hydration enthalpy of $\simeq -640\, \kB T$ (or $-1600$ kJ/mol) \cite{Smith1977} indicates that a cost of $ \simeq 240\, \kB T$, is required to reduce a typical hydration shell of 8 water molecules to one with 5 water molecules, leading us to estimate that a pressure change of roughly $\simeq 1.25\, \si{\giga \pascal}$ would be needed. Our rough calculation obviously overestimates the energetic contribution, which for the first water molecule in a hydration shell is larger than for the eighth, due to steric repulsion, dipole-dipole interactions, and entropic costs. { Numerical studies and X-ray diffraction, that instead consider only the contribution of water molecules in the first shell to the hydration free energy, provide a lower bound of $\simeq 0.67\, \si{\giga \pascal}$ to the energy cost for reducing the Calcium hydration shell \cite{Megyes2004}. The pressure we measure (Fig.~\ref{lowCharge}a) indeed features a non-monotonic dependence on 
	$D$ and the magnitude of the $\simeq 1\, \si{\giga \pascal}$ jump in pressure between $D=9 \,\si{\angstrom}$ and  $D=8 \,\si{\angstrom}$ is well consistent with our estimated range. This shows that} 
	the high stability of the hydration shells can give rise to a competing intermediate-range repulsion \cite{Claesson1986,Shen2021}, confirming early AFM measurements on cement hydrates \cite{Plassard2005} and consistent with gel morphologies obtained in C--S--H coarse-grained simulations and seen in microscopy imaging \cite{Ioannidou2014,Ioannidou2016,Goyal2020}. The finite stability of the bulk-like hydrated structures in increasing confinement provides a fundamental mechanism for this non-monotonic dependence of the nanoscale forces on surface separation, fairly robust to presence of salt and other ions.
	
	At $\sigma = 2$ and $3 e^-\si{\per \nano \meter^2}$ (later hydration stages), the ions are increasingly localized in the $z$ direction and squeezed with their hydration shells against the walls (see ion density profiles in Fig.~\ref{ionProfiles}a,b), becoming unable to maintain full hydration shells in favor of smaller, hemispherical ones. The effect of confinement constraints are very clear at $\sigma = 2 e^-\si{\per \nano \meter^2}$ and $D=\rm 20\,\si{\angstrom}$, where the ion profiles show double peaks, split between ions with two distinct types of hydration shells (Fig. \ref{ionProfiles}c). With further confinement ($D=8\,\si{\angstrom}$), all ions are squeezed against the walls and there is no splitting of the ion density peaks. Finally, for the highest surface charge $\sigma = 3 e^-\si{\per \nano \meter^2}$, the ions stay pressed against the walls even at large separations. With ions localized sufficiently close to the wall, two layers can be accommodated even for the strongest confinement, but the hydration shells are significantly modified by the confining surfaces and are hemispherical for all separations.
	
	\begin{figure*}
		\centering
		\includegraphics[width=\textwidth]{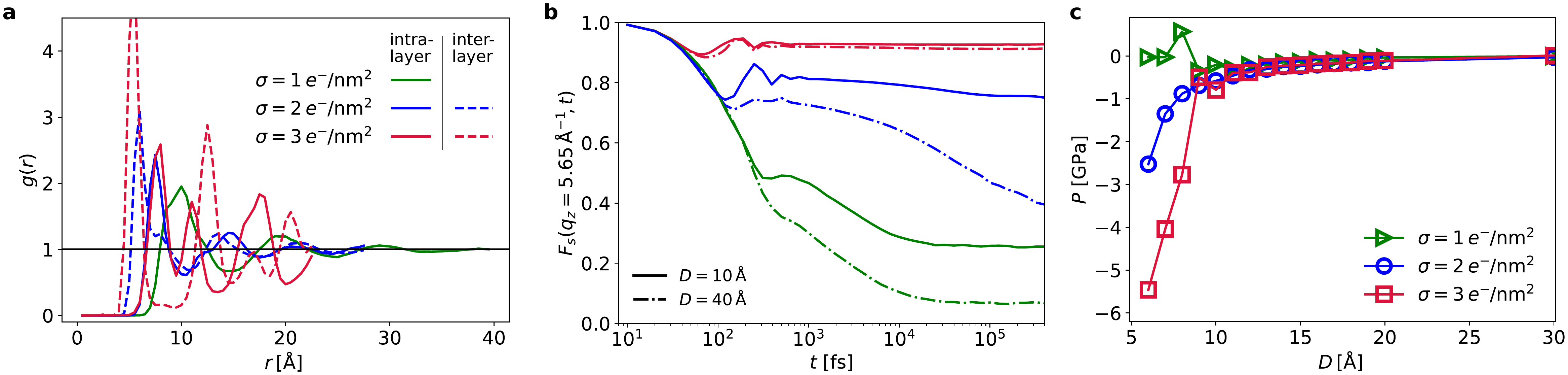}
		\caption{ (a) The $xy$ pair correlation $g(r)$ between ions shows that, as $\sigma$ increases at fixed separation $D= 8\, \si{\angstrom}$, ions become closer together and their positions more correlated. For $\sigma = 1 e^-\si{\per \nano \meter^2}$ the ions are in a single layer equidistant from the confining walls, but at higher $\sigma$ there are two layers, one near each wall. For those, we split the $g(r)$ into the components corresponding to correlations within and between layers, demonstrating a clear spatial organization despite the separation in $z$. { (b) Intermediate scattering functions $F_s(q_z,t)$ for the ions measured for the same $\sigma$ values as in (a) and for two different surface separations $D$. The data demonstrate that the stronger spatial correlations with increasing surface charge density in (a) correspond to increasingly correlated dynamics and more strongly localized ions.} (c) The increasing correlations drive the overall pressure between the confining walls to become increasingly attractive, reaching $P_{\rm min} \simeq -6\, \si{\giga \pascal}$ at $\sigma = 3 e^-\si{\per \nano \meter^2}$.}
		\label{chargeComparison}
	\end{figure*}
	
	At high confinement ($D=\rm 8\, \si{\angstrom}$), 
	the ions in the 
	two layers have distinct but strongly coupled ordering. With layers being defined by the ion position $z_i$, the ion pair correlation $g(r)$ can be separated into intra- and inter-layer contributions, considering ions in the same (intra) or opposite (inter) layers (Fig. \ref{chargeComparison}a). Despite the separation in $z$, the $xy$ positions remain strongly correlated and at high $\sigma$ form a staggered square lattice---the ground state configuration predicted for confined charges in a strong electrostatic coupling regime (defined by high enough surface charge density and strongly confined ions) where ion-ion interactions are included \cite{Samaj2012}. 
{The distance corresponding to first peak of the $g(r)$ decreases with increasing $\sigma$, and for $\sigma = 3 e^-\si{\per \nano \meter^2}$ it is in good agreement with that of the Ca-Ca $g(r)$ recently obtained with x-ray scattering in C--S--H \cite{White2015a,Meral2011}, considering the effect of surface heterogeneities and the surface charge density variation in the real material. 
	
The increase in spatial correlations is associated to strongly correlated dynamics and localization of the ions. The ion intermediate scattering function computed for $q_z = 5.65\, \si{\angstrom}^{-1}$ (roughly corresponding to the peak width of the ion density profiles at the highest $\sigma$) 
is plotted as a function of time $t$ in Fig.~\ref{chargeComparison}b, whereas data for a range of $q_z$ values are provided in SM section \ref{sec:dynamics}. Increasing confinement and surface charge density clearly enhances the dynamical correlations and ion localization, with the effect being particularly dramatic at the highest surface charge. The decrease in mobility and increase in correlation strength} we see with $\sigma$ is coupled to a massive increase (in absolute value) in the net attractive pressure between the two C-S-H surfaces. In fact, at $\sigma = 3 e^-\si{\per \nano \meter^2}$ the pressure minimum is $P_{\rm min} \simeq -6\, \si{\giga \pascal}$ (Fig.~\ref{chargeComparison}c), consistent with atomistic simulations and experiments \cite{Qomi2014,Masoumi2017,Vandamme2009,Geng2017}.
	
	\begin{figure*}
		\centering
		\includegraphics[width=\textwidth]{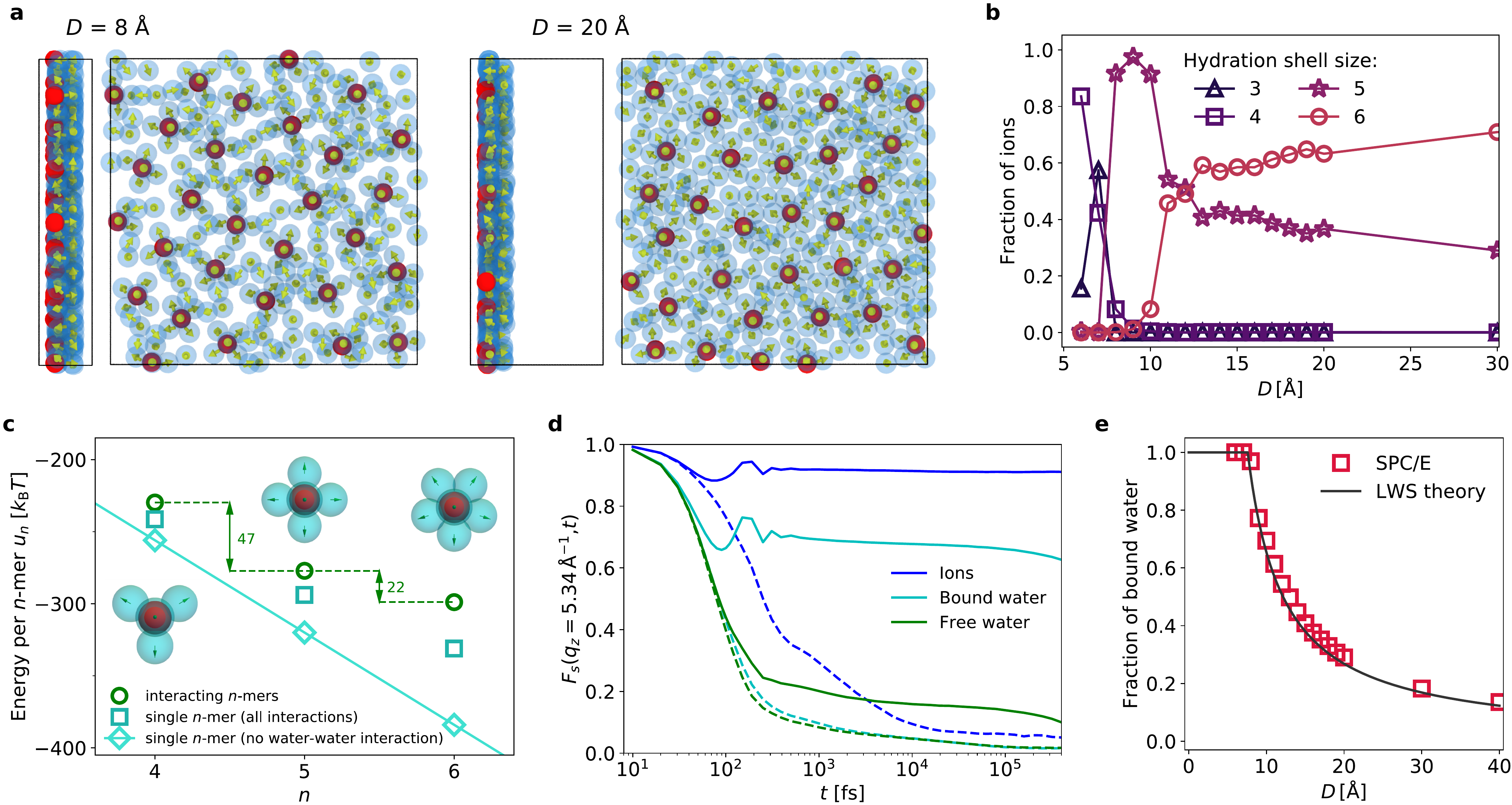}
		\caption{(a) Simulation snapshots and ion-water coordination as a function of surface separation, side and front views. The snapshots show the formation of $n$-mers with an ion and its $n$ water hydration shell. (b) A transition from 3-mers at low separation to a balance of 5- and 6-mers at larger separations is observed. (c) Ground state energy calculations reveal that there is a large gain for a dipole to adsorb on an ion, explaining the formation of these effective objects. {As $n$ grows larger (to 5 or 6), the additional entropic cost becomes sufficient to offset that energetic gain}---with a 6-mer being the largest possible semi-hemispherical object. { (d) Intermediate scattering functions for ions, bound and free water at $D=20\,\si{\angstrom}$ for $\sigma = 3 e^-\si{\per \nano \meter^2}$ (continuum lines) and $\sigma = 1 e^-\si{\per \nano \meter^2}$ (dashed lines). The data show the enhancement of the ion localization and the clead distinction between bound and free water at high surface charge.} (e) The fraction of water in $n$-mers as a function of separation. At small $D$, almost all water is bound up in these $n$-mers, with $n$ limited by water availability. At larger separations, the balance of energetic gain and entropic cost limits $n$-mer size (see Methods and SM section \ref{sec:ionhydration}). These theoretical arguments enable quantitative predictions about the water structure{, via a novel approach based on Locked Water Shells (LWS),} which agree with simulations. }
		\label{nmers}
	\end{figure*}
	
	To further understand the dependence of the pressure, we note that the level of confinement changes the water arrangement around ions (Fig.~\ref{nmers}a), and for $\sigma = 3 \si{\elementarycharge^- \per \nano \meter^2 }$  we reach a balance of 5-mers and 6-mers beyond $D \simeq \SI{15}{\angstrom}$, whereas at even smaller $D$ the surface limits both the number of water molecules and the space available around the ions, leading to a prevalence of 3-mers or 4-mers at the smallest separation $D=\SI{6}{\angstrom}$ (Fig.~\ref{nmers}b).

	Let us now consider that water molecules consist of a spherical particle endowed with a dipole moment, an approach that removes one rotational degree of freedom per water molecule relative to the SPC/E (or TIP4P/2005) water models but allows us to estimate analytically the minimum free energy configurations for the $n$-mers. With this assumption, the energetic gain when a dipole adsorbs to an ion (neglecting any other effect) is $\simeq 64\, k_\mathrm{B}T$ per water molecule (shown by the turquoise line in Fig.~\ref{nmers}c). We can include dipole-dipole interactions within $n$-mers (squares) and then interactions with other hemispherical $n$-mers (circles), to observe that the energetic gain decreases with increasing $n$, but is still more than an order of magnitude higher than $k_\mathrm{B}T$ when going to $n=6$. The minimum energy configurations used in the calculations are sketched in Fig.~\ref{nmers}c and correspond to the shapes observed in simulations for the higher surface charge densities. Taking into account the reduction of water entropy due to the confinement of the molecule and its dipole on the ion, and therefore including finite-temperature in our (so-far) ground-state calculations, one obtains that 5-mers and 6-mers have the same free energy of formation, within a tolerance $\simeq \kB T$ (see SM section \ref{subsec:freeenergy}). This explains the right part of Fig.~\ref{nmers}b, where these two structures appear in commensurate proportions. 
	
	{ The large energetic gain for forming these $n$-mers suggests that they are stable objects, and this is exactly what is observed in the simulations where their lifetimes are found to be longer than the simulation time. By computing the dynamics of the water, we determine that, at the high surface charges where these hemispherical $n$-mers exist, there is now a clear distinction in the behavior of water which is bound in $n$-mers and free water (Fig.~\ref{nmers}d, continuous lines), differently from what happens at low surface charge density (dashed lines).} QENS, DSC, and NMR experiments on cement hydrates indeed provide evidence of distinct populations of unbound and physically bound water molecules emerging during cement hydration \cite{Thomas2001,Bordallo2006,Bohris1998,Fratini2013}. The experimental observation of a bound water fraction increasing with hydration time in cement is perfectly captured here by the dynamical signature of physically bound water at high $\sigma$ (Fig.~\ref{nmers}d) and by the increase in its amount with confinement (Fig.~\ref{nmers}e).
	
	{The effect of confinement can be simply understood by considering that} the free energy gained when a water molecule is adsorbed on a 3-mer to form a 4-mer, or on a 4-mer to form a 5-mer, is energy-dominated and amounts to negative several tens of $\kB T$: it is always extremely favorable to adsorb water molecules on ions from the bulk to increase $n$, at least up to $n=5$. As a consequence, upon increasing the confinement, i.e.\@ when progressively fewer water molecules are available in the nano-slit, all of them are adsorbed on ions. This observation allows us to predict the expected fraction of adsorbed water (Fig.~\ref{nmers}e) and the peaks of the $n$-mers distribution (SM Fig.~\ref{nmersDistributionDM}) for $n=3$, 4 and 5 by assuming all available water is bound in $n$-mers. Using different water models (both SPC/E and TIP4P/2005) does not significantly change these outcomes (see SM section \ref{sec:waterModel}). These findings are also in excellent agreement with fully atomistic simulations of hardened C--S--H (i.e. corresponding to the end of the hydration process) where Ca ions are typically associated just to $3-4$ water molecules \cite{Qomi2014}.  
	
	\begin{figure*}
		\centering
		\includegraphics[width=\textwidth]{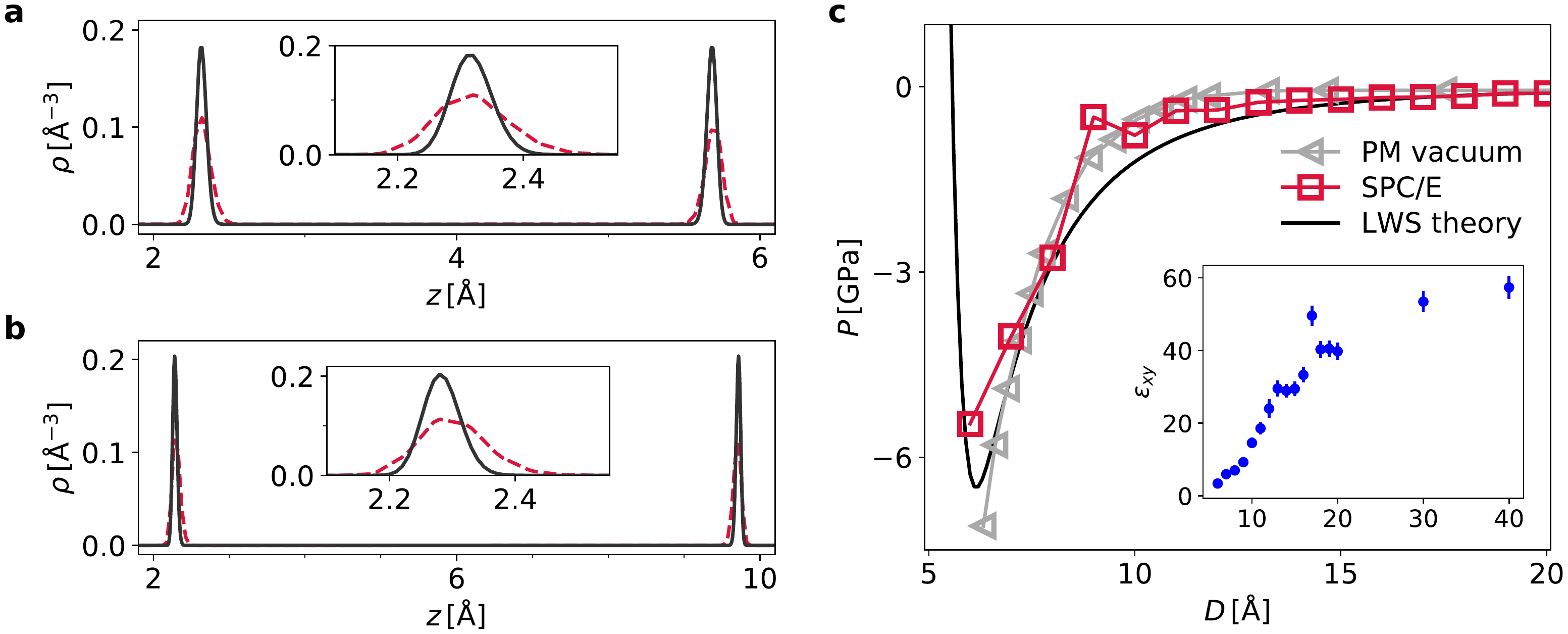}
		\caption{Ion densities can be computed through the theory of strong coupling and correlation holes, considering an $n$-mer as an effective ion and treating appropriately the Lennard-Jones interactions between wall and $n$-mers. These computed density profiles (solid line) are compared to the simulated profiles (dashed line) for (a) $D=\SI{8}{\angstrom}$ and (b) $D=\SI{12}{\angstrom}$. For this quantitative assessment, we compare with simulations that use a dipolar water model to more closely match the analytical calculations. While the theory predicts a somewhat sharper density peak, there is good agreement between the two. (c) From density profiles, one can compute the equation of state, through a generalized version of the contact theorem (see SM section \ref{sec:eos}). This gives a calculated interaction pressure much stronger than earlier theories for implicit water models and close to what is seen in simulations. Due to adsorption of water into $n$-mers at small distances, the effective pressure in the explicit water simulations approaches what one would get for ions in vacuum, for which the rise of pressure at distances below 6 $\si{\angstrom}$ is not visible, since the Lennard Jones contribution has been discarded for the sake of the argument. {The inset shows how the in-plane component of the dielectric tensor is significantly reduced under confinement, consistent with previous studies \cite{Masoumi2019,Fumagalli2018,Schlaich2016} and explaining why water-free results are in surprisingly good agreement with the full explicit water simulations.}}
		\label{theoryPredictions}
	\end{figure*}
	
	It had already been speculated that, because there are so few  water molecules in the interlayers of the hardened C--S--H, the effect of the dielectric properties of water (a fundamental ingredient of PM) on the cohesion should have been significantly reduced \cite{Gmira2004}. We find that the $n$-mers act as stable effective objects and interact in a completely water-depleted environment. The water permittivity tensor $\mathbf{\epsilon}_{r}$ is anisotropic (see SM section \ref{sec:DielectricConstant}), indicating that water dielectric properties cannot generally be recapitulated in a simple constant. Nonetheless, with decreasing $D$, the components of $\mathbf{\epsilon}_{r}$ rapidly approach 1, compared to bulk water permittivity of $\simeq 78$, in line with recent works in other contexts \cite{Schlaich2016,Fumagalli2018} or atomistic simulations of C-S-H \cite{Masoumi2019}. 
	
	To ultimately test this picture and distill its essential ingredients, we develop an analytical theory for the pressure profile beyond DLVO taking advantage of the existence of ``correlation holes'' around each ion \cite{Samaj2018}. The idea is to account for ion-ion correlations by defining a region around each ion in which other ions are prohibited from entering. This is confirmed by the fact that the pair correlation functions $g(r)$ in Figs. \ref{lowCharge} and \ref{chargeComparison} exhibit strongly depleted short-scale features, with $g\simeq 0$. Using this concept, we compute the local effective field $\kappa$ felt by an $n$-mer, due to the presence of all other $n$-mers in a staggered arrangement (see Methods and SM section \ref{sec:eosDensities}). This can be used to calculate the ion density profiles, which match the profiles obtained in simulations, as shown in Fig.~\ref{theoryPredictions}a, b (here we have used only the dipolar interactions for the water to simplify the comparison to the theory, but qualitatively the same behavior is shown in Fig.~\ref{ionProfiles}a, b). While the theory slightly overestimates the density peak heights, this still demonstrates how effective the $n$-mer based theory is at predicting the microscopic details of the simulations. The interactions between $n$-mers and walls yield the pressure between the two surfaces. The Coulombic contribution is given by an analytical pressure equation, which is particularly simple (Eq. \eqref{eq:pressureanalyticalmethods} in Methods). Supplemented with the steric contribution (that becomes relevant only for $D<\SI{6}{\angstrom}$), this yields a total pressure in remarkable agreement with the simulations, see Fig.~\ref{theoryPredictions}c. 
	
{ The fact that our simulations data in the strong coupling regime are so well described by the analytical theory indicates that the underlying ground state is quite resistant to perturbations, because the high surface charge and electrostatic cohesion dominate the overall behavior of the system. Interestingly, when we consider also the relatively good agreement with several aspects of full atomistic simulations and experiments where surface heterogeneities are naturally present, these findings suggest that our simplifying assumption of smooth planar surfaces may be sufficient, in this regime, to capture the overall behavior of the system. This is consistent with the fact that, in strongly correlated systems, the average inter-ion separation expressed by the $g(r)$ is fixed by electroneutrality. As a consequence, in the real material, the same mechanisms revealed here will remain at play, while absolute values of the net cohesive strength will also depend on surface charge density variation, additional surface heterogeneities, and the presence of other ionic species.}
	
	While in our approach water molecules are explicitly accounted for, they are all ``captured'' by ions{---justified by the water dynamics at high confinement (SM section~\ref{sec:dynamics}) {and the consequential reduction in dielectric screening (Fig.~\ref{theoryPredictions}c inset)}}. As a consequence, the typical interaction energies of the system get multiplied by a factor 78, leading to the dramatic increase in the cohesive strength. Ultimately, the effect of the water in terms of dielectric screening is much less important than in bulk conditions, to such an extent that water presence may be ignored altogether. This {``locked water shell'' (LWS)} picture is well illustrated in Fig.~\ref{theoryPredictions}c, showing that the results of simulations with explicit water and from the theory are close to those of PM simulations with a vacuum dielectric constant. The water-ion structuring is a complex function of confinement, surface charge density, and ion species/concentration. However, we can capture the leading effects of the strong electrostatic coupling by the {``locked water'' view}, demonstrating that the net cohesion is due to the ion-water interlocking.
	
	\section{Discussion}
	About one century after the early studies of cement hydration, the properties of C--S--H are still a matter of strong debate. Extensive studies, novel techniques, enhanced characterization and imaging capabilities have greatly improved the knowledge of this complex material, ultimately returning a picture of variability of chemical composition, structural organization and mesoscale morphology. Evidence of this variability has become increasingly clear and is obtained from many sources \cite{Scrivener2015}, with a net increase of data and information available. What is needed, at this point, is to identify the origin of such variability and understand its implications for the material performance, durability and sustainability. 
	
	Our results open the way to do just that. During cement hydration, as C--S--H continously precipitates, ions and water gets progressively confined between increasingly charged surfaces of cement hydrates. We have shown that these two factors, together, change ion-water interlocked structures and their stability, which, in turn, change the net pressure between C--S--H surfaces as hydration proceeds. Our semi-atomistic approach, by including surface charge and ion specificity, captures essential features of cement hydrates detected in experiments, from the ion arrangements to water population and dynamics, to material strength. While it can be extended to include different counterions and mixtures in future studies, here it has provided unique insight into how varying chemical composition that sets the surface charge density changes the water-ion structuring under confinement and hence strength {\it and} shape of the net interactions between cement hydrate nano-particles. { Understanding the origin and the evolution of cement cohesion in terms of fundamental components and mechanisms is key to identifying scientifically guided strategies, such as modifying the ionic composition and increasing cement strength to do more with less, to reduce the greenhouse gas emissions from cement manufacturing.}

	At the nanoscale, the variability of the structural organization of C--S--H in terms of different interlayer separations \cite{Richardson1993,Geng2017} can now be understood via the dependence of the interaction strength, and hence of the energy gain corresponding to different interlayer distances, on the degree of confinement. 	
	However, the evolving shape of the net interactions also has implications for larger length-scales, because it determines the anisotropic growth of cement hydrates aggregates into fibrils, lamellae, and layered mesophases that can self-assemble as C--S--H precipitation proceeds \cite{Ioannidou2014,Ioannidou2016c}. It therefore provides the missing link from the nanoscale to the mesoscale aggregation kinetics and morphological variability of cement hydrates \cite{Richardson1993,Scrivener2015,Goyal2020,Vandamme2009,Geng2017,Bishnoi2009,White2015a,Meral2011,Tennis2000}. By the end of hydration, C--S--H becomes denser and denser, and its nanoscale features (including the interlayer distances and chemical compositions usually described in terms of Ca/Si ratio) play a predominant role in most observations and studies \cite{Geng2017,Masoumi2017,Masoumi2019}. 
Nevertheless, the earlier stage mesoscale morphology controls the development of larger pores and contributes to local stresses in the initial gel network, which have consequences for the long term evolution of the material and its interactions with the environment \cite{Ioannidou2014, Ioannidou2016c, Ioannidou2016,Ioannidou2017,Abuhaikal2018,Zhou2019,Aili2018}. The change in shape of the nanoscale interactions, with competing attraction and repulsion and a striking increase of the attraction strength with surface charge density during hydration, largely controls the morphology of the mesoscale structures that build the gel network and can dramatically steer compressive or tensile stresses as the material progressively densifies and solidifies. These insights shed new light into the physics of cement setting and open new opportunities for scientifically grounded strategies of material design.   
	
	The fundamental understanding of the nature of the electrostatic coupling and the role of ion-water structures has implications beyond cement: a wide range of systems, including biological membranes, soils { and energy storage materials}\cite{Gelbart2000,Levin02,vanDamme2006,Shen2021,Merlet2012}, feature aqueous ionic solutions with both strong Coulombic and confinement effects. While theories such as DLVO work with continuum approaches, we have seen that two discreteness effects interfere with mutual reinforcement: ionic correlation and dielectric destructuring. This calls for a systematic reassessment of electrostatic interactions in strongly confined media, where highly charged objects polarize an ionic atmosphere, in presence of multivalent ions, from clay systems and porous media to water structured interfaces in biological context.
	
	\section{Methods}
	\subsection*{Simulation: model, techniques and parameters}
	All simulations were done using LAMMPS, http://lammps.sandia.gov \cite{Plimpton1995}. We considered a slab geometry which is finite in the $z$ direction and periodic in $x$ and $y$ (the charged surfaces being at $z=0$ and $z=D$), and we ran separate simulations for each value of the surface charge $\sigma$ and $D$. For each set of parameters, exactly 64 calcium counter-ions were included, and the simulation bounds $L_x$ and $L_y$ were adjusted accordingly to ensure overall charge neutrality. The water is treated explicitly, using the rigid SPC/E model \cite{Berendsen1987}. Additional simulations with the TIP4P/2005 model \cite{Abascal2005} were performed for three separations. Results on the microscopic correlations and net pressure are shown in SM section~\ref{sec:waterModel} and closely match those obtained with the SPC/E model. A more computationally expensive water model, as well as polarizability of water molecules, are not expected to play a key role in simulation results (see again SM section~\ref{sec:waterModel}).
	
	The number of water molecules is set using a Grand Canonical Monte Carlo process, discussed in further detail in the next section. To account for the finite size and dispersion interactions of ions, water, and the walls (C-S-H surfaces), we use a Lennard-Jones potential (LJ): 
	\begin{equation}
	U_{\text{LJ}}(r) = 4\epsilon\left[\left(\frac{d}{r}\right)^{12}-\left(\dfrac{d}{r}\right)^{6}\right].
	\end{equation}
	
	In the SPC/E and TIP4P/2005 models, there is one LJ site per water molecule situated at the Oxygen atom. The LJ parameters for the ions are taken from Cygan et al \cite{Cygan2004}. The specific values used for the LJ $d$ and $\epsilon$ are: $d=2.87\,\si{\angstrom}$ and $\epsilon=\SI{0.1}{\kilo\cal/\mol}$ for Ca, $d=3.17\,\si{\angstrom}$ and $\epsilon=\SI{0.155}{\kilo\cal/\mol}$ for SPCE/E water-water, and $d=3.1589\,\si{\angstrom}$ and $\epsilon=\SI{0.162} {\kilo\cal/mol}$ for TIP4P/2005 water. Pairs of different types use the arithmetic mean values of $d$ and $\epsilon$. The wall LJ parameters are the same as those of the SPC/E water. This simplified surface interaction does not consider any heterogeneity or roughness of the surface, which might be relevant for other investigations, but our results indicate that these details are not as important for the properties discussed here.
	
	The final ingredient is the Coulomb forces. Calcium ions are treated as having a point charge of +2e. {The SPC/E and TIP4P/2005 water models have 3 point charges, 2 for hydrogen and one for oxygen, though the exact partial charges and their positions vary between the two models. Detailed information can be found in \cite{Berendsen1987,Abascal2005}.} In order to accurately account for the long-ranged Coulomb forces, we use Ewald summation \cite{Allen1987}. The original formulation is for a 3D periodic system. Yeh and Berkowitz showed that treating the 2D periodic slab system as 3D periodic, with some vacuum space inserted between slabs, is accurate given the addition of a geometry-related correction term to the energy \cite{Yeh1999}. For the slab geometry, this term is:
	\begin{equation}
	E(M,\text{\,slab}) = \dfrac{2\pi}{V} M_{z}^{2}
	\end{equation}
	where $M_{z}$ is the $z$ component of the total dipole moment of the simulation cell.
	
	A few comparisons are made to primitive model (PM) simulations with implicit water. These are performed with exactly the same parameters except there are no water molecules and, instead, the dielectric constant is set to $\varepsilon_r = 78.0$. Effectively, all electrostatic interactions are screened uniformly instead of letting screening effects arise from rearrangement of discrete water molecules. The resulting disparity is due to the fact that this is an insufficient representation of the effects of water in this system. Comparison is also made with PM calculations in vacuum (i.e. with $\varepsilon_r = 1$); this gives credence to the ``dry water'' view discussed in the text for the strong-coupling regime.
	
	In order to make direct comparisons with the analytical predictions that use only the dipolar term in the multipole expansion for water, some simulations were performed with purely dipolar interactions for the water molecules (namely Fig.~\ref{theoryPredictions}a,b, \ref{fig:orientation}, and \ref{nmersDistributionDM}). In these cases, the water Coulomb interactions were computed using a point dipole with a moment $m = 0.37\,e\si{\angstrom}= 1.8$\,D, selected so that the dielectric constant matches that of water in bulk (room temperature, pressure) conditions. 
	
	\subsubsection*{Water Density}
	Water in confined geometries (especially in the presence of charges) can have a density that is different from bulk conditions. This density is highly dependent on the level of confinement and the strength of the electric fields in the system because of their effect on water structure, so it is a function of $D$ and $\sigma$. To select the number of water molecules for our simulations, we first performed Grand Canonical Monte Carlo (GCMC) based on the chemical potential of bulk water (room temperature, density). Simulations in bulk conditions showed that a chemical potential of $\mu=\SI{-8.8}{\kilo\cal/\mol}$ gave the correct water density. 
	
	This chemical potential can be maintained using the Metropolis method by attempting insertions/deletions with equal probability, and accepting them with the following probabilities \cite{Allen1987}:
	\begin{equation}
	p_{\text{ins}}=\min\left(1,\dfrac{V}{\Lambda^3(N+1)}e^{\beta(\mu-\Delta U)}\right)
	\end{equation}
	\begin{equation}
	p_{\text{del}}=\min\left(1,\dfrac{\Lambda^3N}{V}e^{-\beta(\mu-\Delta U)}\right)
	\end{equation}
	Here $\beta=1/(\kB T)$, $\Lambda$ is the thermal De Broglie wavelength and $\Delta U$ is the internal energy change upon attempting a particle insertion or deletion. Using a Monte Carlo simulation, with random insertions and deletions, the Grand Canonical ensemble is sampled and, in equilibrium, the chemical potential is maintained with water density fluctuating around a mean value.
	
	The GCMC process alone is very slow to converge to the final density as it only considers single molecule moves. To speed up the convergence, we combined this with Molecular Dynamics (MD), shown previously to significantly decrease the convergence time \cite{Carrier2014}, with 1000 GCMC exchanges (insertions, deletions) attempted every 1000 MD steps with a time step of $1\,\si{\femto\second}$. Starting with $0$ water molecules, the GCMC and MD simulations were run until equilibrium was reached {(when energy, pressure, and water density no longer changed with time).} With this process, the equilibrium density in confinement was found within $3\cdot10^6$ MD steps for the $\sigma=3\,\si{\elementarycharge/\nano\meter}^2$ simulations.
	
	In the low $\sigma$ simulations (with larger system size), the convergence was found to be very slow due to the larger number of water molecule insertions required. After $3\cdot10^6$ MD steps and GCMC exchanges, it remained unclear if the water density had reached the equilibrium value. In this situation, further simulations were performed using initial configurations with randomly placed water molecules at bulk density ($\rho = 1\,$g/cm$^3$). As the initial water density in these configurations was closer to the final one, fewer GCMC exchanges were required to reach an equilibrium value and this significantly reduced the GCMC simulation time required to converge to the equilibrium density. In addition, we were able to observe convergence towards the final water density from different initial conditions, making it clear that the chosen values were appropriate.
	
{	\subsubsection*{Analysis of pressure and spatio-temporal correlations} }
	After determining the equilibrium number of water molecules for a system with given parameters, { we have confirmed that all systems have reached equilibrium by checking that there was no drift in the energy and pressure over time, that time correlations had substantially decayed and/or there was no aging sign in two-time correlation functions.} 
	We have then generated trajectories in the NVT ensemble, using molecular dynamics and the velocity Verlet algorithm with an integration step of $1\,\si{\femto\second}$.
	All simulations data discussed here have been averaged over $10^{6}$ MD steps. {To ensure sufficient thermodynamic sampling in the cases of high confinement (where dynamics are substantially slower), additional simulations using 10 independent sets of random initial conditions were performed for all samples with $D\leq12\,\rm \si{\angstrom}$. Both initial ion velocities and the random number seed for the GCMC process were varied, resulting in slight differences in the number of water molecules. In all cases, sample-to-sample fluctuations were smaller than the symbol size in our plots, with a maximal standard deviation in pressure (at $D=6\,\rm \si{\angstrom}$) of $\rm s_{max}<.05\,GPa$.} From the particle trajectories, the pressure and microscopic correlations were computed. The pressure was calculated as the time average of the total force exerted on one of the C-S-H surfaces, which fluctuates around a mean value in equilibrium. {This was computed for each value of $D$, and the pressure profile was obtained from subtracting the large distance ($D=40\,\si{\angstrom}$) value of the pressure.}
	
	To investigate the microscopic origins of this force, we also study the spatial and dynamical correlations that arise in the ions and water. There are theoretical predictions that ions have strongly correlated positions and may even form 2D crystals (see \cite{Samaj2012} and SM section \ref{sec:coupling}). To quantify the extent to which this holds, we calculated the pair correlation function $g(r)$ of ions in the $xy$ plane, i.e. parallel to the C-S-H surfaces, defined as:
	\begin{equation}
	g(r) = \frac{L_x L_y}{2\pi r\Delta r N_{\rm ion}^2 } \left\langle \sum_{i} \sum_{j\neq i} \mathcal{H}\left(\frac{\Delta r}{2} - \abs{r-r_{ij}}\right)    \right\rangle
	\label{eq:gr}
	\end{equation}
	where $r_{ij}$ is the distance between the two ions in the $xy$ plane, $\Delta r$ is a binning distance, and $\mathcal{H}$ is the Heavyside function. This function was also calculated for specific groups of ions, determined by their $z$ position, which simply involved modifying $N_{\rm ion}$ and the $\sum$ bounds appropriately. 
	
	{ 
	In addition to static spatial correlations, significant dynamical correlations also arise. These have been studied by computing the self-intermediate scattering function {\cite{Hansen2006}}, which quantifies the time correlations of ion (or water molecule) displacements. In particular, we analyzed the dynamics in the direction normal to the surface, $z$:
	
	\begin{equation}
	F_s(q_z,t) = \frac{1}{N} \left\langle \sum_{j=1}^{N} \mathrm{e}^{iq_z(z_j(t)-z_j(0))}      \right\rangle \,,
	\label{eq:Fscatt}
	\end{equation}
	where the $q_z$ values are determined by the system dimension $D$ and the boundary conditions. The $q_z$ value sets the length scale of displacements that contribute most to $F_s(q_z,t)$, and due to boundary conditions the lowest allowable value is $q_z = 2\pi/D$. We consider all multiples of this value up to $q_z=10\, \si{\angstrom}^{-1}$, corresponding to sub-angstrom length scales. To separate the contributions from different groups of ions or water molecules, we modify the $N$ and the bounds of the $\sum$.}
	\\
	\subsection*{Theory of the electrostatic coupling for high surface charge}
	
	A useful and commonly used tool to describe charged solutions at equilibrium, the Poisson-Boltzmann approximation (at the root of the electrostatic contribution to DLVO theory), is a mean-field approach: ions are treated as a charged cloud, whose density depends on the \textit{average} electrostatic potential.
	Spatial correlations between ions, and more generally discreteness effects are discarded, since the charged cloud is viewed as continuum. It can be shown that within Poisson-Boltzmann theory, the force between two like-charged surfaces is always repulsive \cite{Israelachvili2011}. 
	Accounting for the discrete nature of ions explicitly, 
	attraction may set in under strong enough Coulombic coupling, as we next explain. A fingerprint of the mechanism behind attraction may be found in the $xy$ staggering of ionic patterns between one wall and the other (staggered peaks between the intra-layer $g(r)$ and the inter-layer $g(r)$, as visible in Fig.~\ref{chargeComparison}a).
	
	For a salt-free system such as the one under study here, Poisson-Boltzmann theory is a trustworthy approximation under conditions of weak electrostatic coupling \cite{Moreira2001,Samaj2018}. On the other hand, mean-field fails and attraction can take place due to ionic correlations when coupling becomes strong. Electrostatic coupling is quantified through the parameter
	\begin{equation}
	\Xi = \frac{q^2 \lB}{\mu_\mathrm{GC}} = 2 \pi q^3 \lB^2 \sigma \, ,
	\end{equation}
	where $q$ is the counterions' valence, $\sigma$ is the absolute value of the surface charge density of the walls divided by the elementary charge $e$,  $\lB={e^2}/({4 \pi \varepsilon_0 \varepsilon_r \kB T})$ (the Bjerrum length) is the distance between two elementary charges such that their repulsive potential energy is $\kB T$, and $\mu_\mathrm{GC} = (2 \pi \lB q \sigma)^{-1}$ (the Gouy-Chapman length) is the distance of a charge $qe$ from the wall such that its attractive potential energy is $\kB T$. In practice, $q^2 \lB$ quantifies the strength of electrostatic repulsion between counterions, while $\mu_\mathrm{GC}$ quantifies how close to the wall ions tend to stay. When $\Xi$ is small (weak coupling), ions can come relatively close to each other and populate the region within a distance $\mu_\mathrm{GC}$ from the wall without need to form any ordered structure. Conversely, when $\Xi$ is large (strong coupling), ions lie very close to the walls, but feel a strong mutual repulsion, so they need to form ordered planar structures to minimize their energy. 
{	
\subsubsection*{Ion density and pressure}
At strong coupling (the relevant situation here), an ion on either wall is subjected to an effective electric field written for convenience as $\kappa / (\beta q e \mu_\mathrm{GC}) = \kappa \sigma / (2 \varepsilon_0 \varepsilon_r)$. 
The quantity $\kappa$ can be viewed as the dimensionless local electric field acting onto an ion, with $0 < \kappa <1$. It depends on wall-wall distance $D$ and stems from electrostatic correlations. 
For large $D$, ions that are located in the vicinity of one of the two walls mostly feel the field due to this wall, the other one being screened by the remaining ions: this means $\kappa\simeq 1$.
In the opposite limit, when $D$ becomes smaller than the typical ion-ion distance, an ion feels the field created by both walls, in opposite directions, while the contribution due to other ions is subdominant. In this small-$D$ regime where all ions essentially lie in the same plane, $\kappa$  has to vanish. An analytical expression of $\kappa$ as a function of $D$ has been computed for point-like particles and hard walls in \cite{Samaj2018}. 

In our simulations, the presence of a soft wall potential does not modify much the picture, except for the fact that $D$ must be replaced by $D_{\text{eff}}=D-2 z_c$, where $z_c$ is the equilibrium distance of an $n$-mer (a Ca ion dressed with $n$ water molecules) from the closest wall.
Taking this into account, we find the expression for the effective field given in Eq.~(\ref{eq:kappa}) and plotted in Fig.~S4b. 

The average ion ($n$-mer) density is then computed as per Eq.~(\ref{eq:rho}), which
can be obtained by deriving the partition function of an ionic system within soft walls, similarly to what is done in \cite{Samaj2018}. 
This is the density that is plotted in Fig.~5a,b.

The pressure shown in Fig.~5c is eventually computed from the ion density through Eq.~(\ref{eq:pressurecontacttheorem}), which is an extension of the contact theorem, a known exact result. This amounts to splitting the pressure into an electrostatic contribution and a kinetic one, due to interactions with the walls. 
Before and close to its minimum point, at very small distances, the pressure profile turns out to be dominated by Lennard-Jones repulsion and can be approximated by Eq.~(S10). After the minimum, the pressure increases in agreement with the analytical hard-wall prediction
\begin{equation}
\label{eq:pressureanalyticalmethods}
P(D) = 2 \pi \lB \sigma^2 \kB T \left[ -1+ \kappa(\Deff)\left( \frac{1+e^{\kappa(\Deff) \frac{\Deff}{\mu_\mathrm{GC}}} }{ 1-e^{\kappa(\Deff) \frac{\Deff}{\mu_\mathrm{GC}} }}  \right) \right] .
\end{equation}
This is the relevant regime for our study, exhibiting negative $P$. Note that the pressure prefactor scales as the square of the surface charge and, most importantly, as the inverse of the dielectric permittivity: within our locked water theory, pressure is enhanced by almost two order of magnitudes, compared to the primitive model. 

More details are given in Sec.~\ref{sec:eos} of the SM.
}	
	\subsubsection*{Energy of $n$-mers}
	The formation of $n$-mers due to hydration of Ca ions is an energetically driven process for $n<5$, while entropy comes into play for larger $n$. The minimum energy of an $n$-mer was computed (within the dipolar model, more convenient to this end) through a simulated annealing procedure, where an ion was considered fixed next to a wall and energy was minimized with respect to the positions and orientations of $n$ water molecules. Due to the presence of the wall, the volume available to water molecules is restricted to a hemisphere. Considering the minimum-energy configuration of an $n$-mer, the formation energy $u_n$ per $n$-mer is then refined, accounting for the presence of the (infinitely many) other $n$-mers, lying in the same plane. Due to the long range nature of Coulombic interactions, it is important to keep track of all neighbors. $u_n$ turned out to be $\lesssim -\,50\,n\,\kB T$ for $n\le6$ (see Fig.~\ref{nmers}c), while $n\ge7$ was not considered, as the minimum-energy configurations are such that no more than 6 water molecules can fit in one hydration shell. More details are given in SM section \ref{subsec:energy}.

	\subsubsection*{Free energy of $n$-mers}
	The free energy $\Delta F_{n \rightarrow n+1}$ of formation of an ($n+1$)-mer, from the absoption of a water molecule on an $n$-mer, can be assessed using the previously computed values of $u_n$, the chemical potential $\mu$ of bulk water used in simulations, and estimating the associated entropy change. Entropy decreases due to increased confinement of water molecules: this can be quantified by single-particle calculations of the volume available to every water molecule when it is bound to an $n$-mer, as a function of $n$. For $n=3$ and $4$, we have $\Delta F_{n \rightarrow n+1} < -30 \kB T$, while $\Delta F_{5 \rightarrow 6} \simeq 0$ within a few $\kB T$. Details can be found in SM section \ref{subsec:freeenergy}. Since adsorbing water molecules to increase $n$ is extremely favorable, at least up to $n=5$, one can assume all molecules to be bound at strong confinement and predict quantitatively distances $D_n$ at which all ions are bound to exactly $n$ water molecules: we find
	\begin{equation}
	\label{eq:limitedres}
	\rho_{\text{w}}\, D_n \, = \,2\,n\, \frac{\sigma}{q} \,,
	\end{equation}
	where $\rho_{\text{w}}D$ is the average number of water molecules per unit surface at distance $D$.
	
	These simple arguments are in excellent agreement with  numerical simulations. Our calculations explain the fraction of bound water (Fig.~\ref{nmers}e) observed in simulations as a function of $D$, and the fraction of 5-mers to 6-mers in simulations with a dipolar approximation for the Coulomb interactions (Fig.~\ref{nmersDistributionDM}). More details on this limited resources argument are given in SM section \ref{sec:limitedresources}.
	
	\section{Acknowledgements}
\textbf{Funding:} This work has received funding from the European Union's Horizon 2020 research and innovation programme under the Marie Sk\l{}odowska-Curie grant agreement 674979-NANOTRANS.
AG and EDG acknowledge the NIST PREP Gaithersburg Program (70NANB18H151) and Georgetown University for support.
{
	\textbf{Competing Interests:} The authors declare that they have no competing interests.
	\textbf{Data Availability:} All data needed to evaluate the conclusions in the paper are present in the paper and/or the Supplementary Materials.
	\textbf{Author Contributions:} A.G. performed all numerical simulations and data analysis. I. P. performed the analytical calculations. R.J.M.P, E.T. and E.D.G designed the research. All authors provided critical feedback and helped shape the research, analysis, and write the manuscript.
	\\
	(*) Indicates A.G. and I.P. contributed equally
	\\
	($^+$) Indicates A.G. and E.D.G. are corresponding authors
	
	\section{Supplementary Materials}
	The Supplementary Materials contains figures S1-S9 and sections S1-S4 discussing ion and water dynamics, the strong coupling theory, properties of water in simulation, and a historical overview of cement.
}

\pagebreak

\include{SupplMat_noColor}

\end{document}

%% file: SupplMat_noColor.tex
\setcounter{section}{0}
\setcounter{figure}{0}
\setcounter{page}{1}
\setcounter{tocdepth}{4}
\setcounter{secnumdepth}{2}
\setlength{\parskip}{0.2\baselineskip}

\renewcommand\thesection{S\arabic{section}}
\settowidth\cftsecnumwidth{S10000000}
\renewcommand\thesubsection{S\arabic{section}.\arabic{subsection}}
\settowidth\cftsubsecnumwidth{S100000000.000}
\renewcommand\thefigure{S\arabic{figure}}
\renewcommand\theequation{S\arabic{equation}}

\newcommand{\pard}[2]{\frac{\partial #1}{\partial #2}}
\newcommand{\pardd}[2]{\frac{\partial ^2 {#1}}{\partial {#2} ^2}}
\newcommand{\parddd}[2]{\frac{\partial ^3 {#1}}{\partial {#2} ^3}}
\newcommand{\pardddd}[2]{\frac{\partial ^4 {#1}}{\partial {#2} ^4}}
\newcommand*\diff{\mathop{}\!\mathrm{d}}
\newcommand{\eff}{\mathrm{eff}}
\newcommand{\LJeff}{\mathrm{LJ,eff}}
\newcommand{\LJcd}{\mathrm{LJ,cd}}
\newcommand{\LJwall}{\mathrm{LJ,w}}

\newcommand{\lBwat}{l_\mathrm{B}^{\mathrm{wat}}}
\newcommand{\muGCz}{\mu_\mathrm{0}}

\graphicspath{{figures-SM/}}

\onecolumngrid
\titleformat{\section}{\huge\bfseries}{\thesection}{1em}{}
\titleformat{\subsection}{\large\bfseries}{\thesubsection}{1em}{}
\titleformat{\subsubsection}{\bfseries}{}{1em}{}

\section*{Supplementary Material}
\subsubsection*{A. Goyal, I. Palaia, K. Ioannidou, F. Ulm, H. van Damme, R. Pellenq, E. Trizac, E. Del Gado}

This Supplementary Material contains the discussion of the ion-water dynamics for different surface charge density, based on the numerical simulations. We also provide further details on the construction of the correlation hole theory and discuss the properties of water in the simulations.  Finally we conclude with an excursus giving a historical overview of cement. 

\titleformat{\section}{\large\bfseries}{\thesection}{1em}{}
\section{Ion and water dynamics}
\label{sec:dynamics}

{
In this section, we expand on the dynamics data shown in Figs.~\ref{chargeComparison}b and \ref{nmers}d. The plots in the main paper showed data for a single $q=q_z$ value, picked to show the trends in dynamics with $\sigma$ and $D$, and here we include the data for all $q$ values for the different $\sigma$ and $D$ considered (with evenly spaced values of $q$ ranging from $q=2\pi/D$ to $q=\SI{10}{\angstrom^{-1}}$). We also take this opportunity to explain in more detail the comparisons with experimental measurements.
} 

\begin{figure*}
	\centering
	\includegraphics[width=\textwidth]{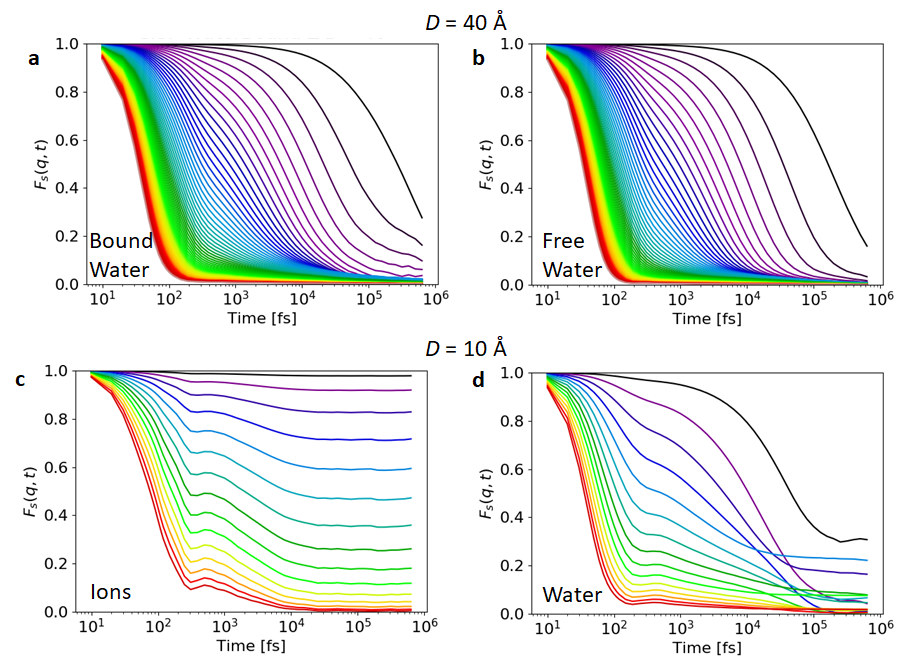}
	\caption{Self intermediate scattering function at $\sigma = 1 \si{\elementarycharge^- \per \nano \meter^2 }$ in the $z$ direction (normal to surface plane), for ions or water. Color indicates $q$ from $q=2\pi/D$ for black to $q=\SI{10}{\angstrom^{-1}}$ for red. Results for bound (a) and free (b) water at $D=40\,\si{\angstrom}$, classified by whether they are bound to an ion at $t=0$, exhibit significant decays in correlation over simulation times: though slightly slower for bound water than free water. The water dynamics diverge from that of the ions due to the finite ion-water bond lifetime. However, upon increasing confinement, the situation changes. Ions (c) at $D=10\,\si{\angstrom}$ exhibit plateaus due to the formation of layers and limited mobility. This partially extends to the water (d) which exhibits much slower decays compared to the water at $D=40\,\si{\angstrom}$ (ion correlations are relatively unaffected by changes in $D$). Additionally, we find that for specific $q$ values ($4\pi/D$, $8\pi/D$) the correlations exhibit a larger decrease, suggesting that these length scales correspond to specific distances at which elementary rearrangements of the water molecules can occur in confinement. Note that such $q$-dependent features are erased at larger separations where water mobility is less limited.}
	\label{dynamics_sig1}
\end{figure*}
{
At large separations, most of the water is not associated with ions. For water that is bound in hydration shells at time $t=0$ (plotted in Fig. \ref{dynamics_sig1}a), we observe a complete decay in correlations over the simulation--albeit at a slightly slower rate than with the free water (Fig. \ref{dynamics_sig1}b). While energetically favorable, Calcium hydration shells are highly dynamic with residence times $t_{res} \simeq 7 \cdot 10^5\,\si{\femto \second}$ \cite{Koneshan1998}, and we observe significant decay in $F_s(q_z,t)$, even for bound water, indicating that the water can move back and forth between free and bound states and does not stay closely attached to the ions, which are instead clearly localized.
In confinement, the ions become highly localized in the $z$ direction, as demonstrated by plateaus in $F_s(q_z,t)$ (Fig. \ref{dynamics_sig1}c). Due to the formation of hydration shells and geometric/packing constraints in confinement, this localization extends to the water, as seen by the relatively slow decays in Fig.~\ref{dynamics_sig1}d (quite slower when compared to the dynamics shown in Fig.~\ref{dynamics_sig1}b). Water-ion bond lifetimes measured from bond correlations are found on the order of $O(10^5 \rm fs)$, indicating that correlations in water dynamics should still decrease faster then for ions, as we indeed observe. In addition, we find that for specific $q$ values ($4\pi/D$, $8\pi/D$) the correlations exhibit a larger decrease, suggesting that these length scales correspond to specific distances at which elementary rearrangements of the water molecules can occur in confinement. 
}

\begin{figure*}
	\centering
	\includegraphics[width=\textwidth]{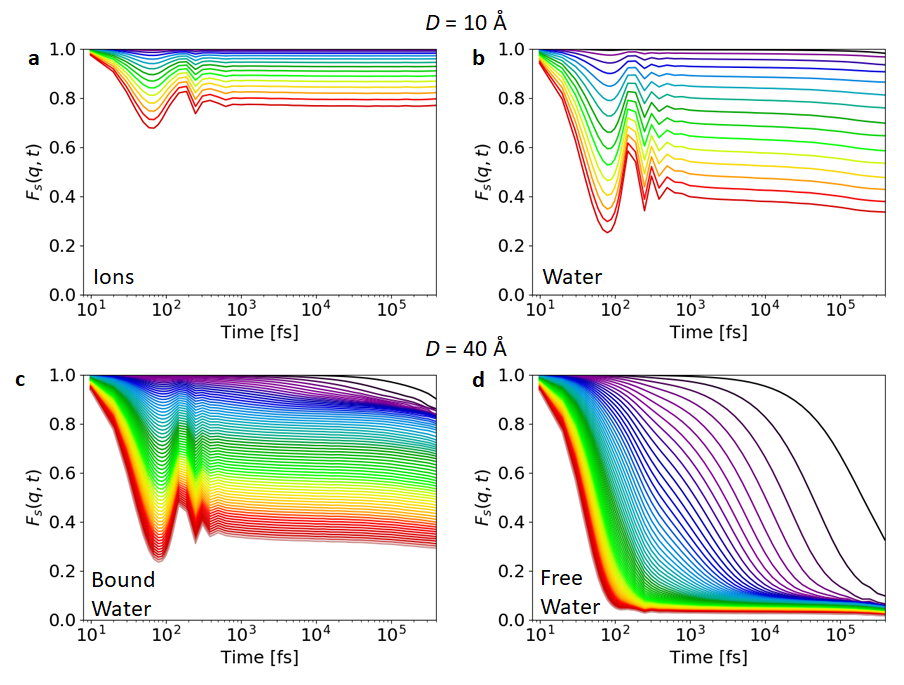}
	\caption{Self intermediate scattering function at $\sigma = 3 \si{\elementarycharge^- \per \nano \meter^2 }$ in the $z$ direction (normal to surface plane), for ions or water. Color indicates $q$ from $q=2\pi/D$ for black to $q=\SI{10}{\angstrom^{-1}}$ for red. (a) Ions in confinement exhibit strong localization near the surfaces. Notably, this does not change when considering larger separations. (b) Same quantity for water. Its dynamics are highly coupled to ion dynamics in confinement. At larger separations, water is split into two populations. The bound water (c), which is close to ions at $t=0$, follows the ion dynamics, while free water (d) is much more mobile. The dynamical signature of the ions appearing in this bound water demonstrates the stability of ion-water structure and its persistence at larger separations. }
	\label{dynamics}
\end{figure*}

Upon increasing surface charge to $\sigma = 3 \si{\elementarycharge^- \per \nano \meter^2 }$, which would correspond to the end of hydration, we discover a drastic change in the dynamics. $F_s(q_z,t)$ for the ions (plotted in Fig. \ref{dynamics}a) exhibits even stronger localization with very high plateaus in the correlation. As this effect persists at $q=\SI{10}{\angstrom^{-1}}$, this localization holds for length scales smaller than an angstrom, and we infer that this signal corresponds to the localization of the ions near the surfaces, consistent with the density profiles averaged over time shown in Fig. \ref{ionProfiles}. The plot shown is for a separation of $D=\SI{10}{\angstrom}$, but $F_s(q_z,t)$ for the ions is very similar up to $D=\SI{40}{\angstrom}$---the highest separation simulated.

Starting at low separation (Fig. \ref{dynamics}b), we see that the water behavior closely follows the ion dynamics. The correlations drops off to a lower valued plateau, meaning the localization is not quite as strong as for the ions, but it is clearly there for the water as well. The oscillations in Fig.~\ref{dynamics}b also mirror those exhibited by the ions, showing how strongly the water dynamics are coupled to those of the ions. These strong and long-lasting dynamical correlations are evidence of the formation of strongly correlated ion-water assemblies. While similar structures are observed at lower surface charges, the residence time of water molecules in the hydration shells is much lower, and thus the water correlations are not as long-lived. Instead, at high $\sigma$, the ion-water assemblies persist through the simulation time and the bound water remains highly localized even for $t>10^5 \, \si{\femto \second}$.

The same picture persists even at larger separations (Fig. \ref{dynamics}c). However, while at lower separations most of the water is bound to ions, at $D=\SI{40}{\angstrom}$ we observe drastically different behavior for bound and free water. The bound water, i.e. the water in the ion hydration shells, behaves exactly the same at $D=\SI{40}{\angstrom}$ as at $D=\SI{10}{\angstrom}$. This water is strongly coupled to the ions and they move (or rather do not move) in unison. Instead, the dynamics of free water (Fig. \ref{dynamics}d) is substantially uncorrelated from the ions. This ``free'' water is, of course, still confined by the C-S-H walls and exhibits relaxation times comparable to those measured via experiments on water in cement pores \cite{Zhang2008}. However, the fact that there is no localization in the free water, indicates that the dynamical behavior of the bound water is largely determined by electrostatic interactions with ions and thermodynamics of $n$-mers rather than the confinement effects.

{
	This clear separation of dynamics has been documented extensively in experiments on hydrating cement. Quasi-elastic neutron scattering (QENS) and nuclear magnetic resonance (NMR) experiments demonstrate that, as hydration progresses and C-S-H stoichiometry evolves while it progressively aggregates into a porous structure, an increasing fraction of water becomes significantly slower and is considered as physically bound \cite{Thomas2001,Bordallo2006,Bohris1998}. Hence in hydrating cement there exist three populations of water: chemically bound, constrained (i.e. physically bound), and unbound. While our model does not resolve the chemically bound water, the characterization we provide of the “bound” and “free” water is, in fact, in good agreement with the QENS and NMR experimental characterization of constrained and unbound water. The intermediate scattering functions we compute are, in fact, directly related to the time correlations of the scattering intensity measured in QENS experiments (with the caveat that here we are computing the signal from the incoherent scattering, but the same signature would be seen in the coherent one). 
	
	Not only do our data confirm the experimental observations on the physically bound water, but they also provide the insight that such water is the one corresponding to the hydrated (or partially hydrated) ions, which constitute our ion-water interlocked structures.  Our simulations show that this effect is dependent on the surface charge density $\sigma$, and at low $\sigma$ the water bound to ions is dynamically similar to the free water (Fig.~\ref{dynamics_sig1}). Additionally, at fixed $\sigma$, an increase in confinement also raises the fraction of bound water (Fig.~\ref{nmers}e). Both effects are consistent with the experimentally observed increase in physically bound water as a function of hydration time in cement. Changes in the bound water fraction due to C-S-H stoichiometry, including the Ca/Si ratio, can be understood through the effect those changes have on surface charge densities and interlayer distances (i.e. confinement), which in our picture indeed control the ion-water structures. The combined experimental/modeling work of Geng et al \cite{Geng2017}, for example, shows that the Ca/Si ratio controls the interlayer spacing of C-S-H. As a side note, they also report that decreasing interlayer spacing and an increase in calcium ions is responsible for a rise in the bulk modulus of C-S-H, which is consistent with our results on the cohesion strength.
}
\section{Strong coupling theory for high surface charge density}
\label{sec:coupling}
At strong confinement, water destructuring heavily alters its electrostatic screening properties. This is due to the presence of solvation shells around Ca$^{2+}$ ions: as shown in Fig. \ref{ionProfiles} for $\sigma = 3 \si{\elementarycharge^- \per \nano \meter^2 } $, indeed, ions tend to stick to the walls. We observed that up to a certain distance $D$ between the walls, practically all available water molecules are used by the system to hydrate Ca$^{2+}$ ions and none is free to move. Figure \ref{nmers}e shows that more than 70\% of water molecules are bound to an ion for separations shorter than \SI{10}{\angstrom}. 
The hydration of ions results in a huge energy decrease. Albeit associated with a conspicuous entropy reduction, due to positional and orientational localization of water molecules, they are generally favorable in terms of free energy, as we will show in SM section \ref{subsec:freeenergy}. This is why, at strong confinement no water molecule is free to move between the two double layers and therefore electrostatic interactions between hydrated ions are not screened. Hydrated, or ``dressed'' ions ($n$-mers) behave as effective charged objects, interacting in vacuum: this is referred to in the main paper and in the following as the ``locked water'' picture. 

As a consequence, the relevant Bjerrum length is not $\lB = \frac{\beta e^2}{4 \pi \varepsilon_0 \varepsilon_{r}}$ (with $\beta^{-1}=k_BT$), as in bulk water: it is rather close to $\lBz = \frac{\beta e^2}{4 \pi \varepsilon_0}$, the Bjerrum length in vacuum, that is a factor $\varepsilon_r \simeq 78$ times larger. The effect of this is twofold: 1) the minimum pressure predicted by the contact theorem \cite{Andelman2010} increases (in absolute value) by a factor 78 from $2\pi \lB \sigma^2$ to around $2\pi \lBz \sigma^2$, due to unscreened Coulombic interaction; 2) the coupling parameter $\Xi$, defined in the Methods, increases from $2 \pi q^3 \lB^2 \sigma \simeq 75$ to around $2 \pi q^3 \lBz^2 \sigma \simeq 480\,000$ for $\sigma = 3 \si{\elementarycharge^- \per \nano \meter^2 }$, thus amply justifying the use of strong coupling theory in the following. These two factors determine a pressure two orders of magnitude higher, in absolute value, than what predicted by primitive models treating water as a dielectric continuum (see SM section \ref{sec:eos}). Our simplification with a drastic decrease of the dielectric permittivity of water, due to confinement, is backed up by a numerical estimate of $\varepsilon_r$ based on simulations and the related observations presented in SM section \ref{sec:DielectricConstant}.

Staggered Wigner crystals have been observed in the literature \cite{Moreira2002,Boroudjerdi2005} for coupling higher than $31\,000$. Our large value of the coupling parameter explains then the perfect staggered square crystal observed in Figure \ref{nmers} for high $\sigma$ at distance $D=\SI{8}{\angstrom}$, referred to in the literature \cite{GoPe96,Samaj2012} as phase III. Due to Lennard-Jones repulsion, the effective distance between the two ionic layers is indeed $D_{\mathrm{eff}}\simeq \SI{3.5}{\angstrom}$: this corresponds to a dimensionless distance $\eta=D_{\mathrm{eff}}\sqrt{\frac{\sigma}{q}}\simeq0.43 $, at which phase III is expected \cite{GoPe96,Samaj2012}. In addition, only such a strong coupling can explain the ionic density in Figure \ref{theoryPredictions}a, which is strongly peaked close to the walls and vanishes in the whole central region (strong coupling theory does not forbid a uniform ion density, but such a profile appears only at distances $D_{\mathrm{eff}}$ ten times smaller).

\subsection{Ion hydration}
\label{sec:ionhydration}

\subsubsection{Energy}
\label{subsec:energy}

We focus here on the mechanism by which water molecules tend to bind to ions and study the energy gain associated to the formation of an $n$-mer. In simulations, cations tend to lie on parallel planes, at a distance from the closer wall given by the balance between Lennard-Jones interaction and electrostatic attraction. In doing so, they place themselves as close as possible to the walls, so that they expose only a half of their surface for binding with water. This allows us to consider the hydration shell around each ion a hemisphere, around which no more than 6 water molecules can fit. A good estimate of the energy scale at stake is the single dipole-ion interaction energy; in purely electrostatic terms, it can be estimated, in units of $\kB T$, to $ -\frac{q \delta \lBz}{\sigma_{\LJeff}^2}\simeq -70$, where $e\delta=0.375\,e\si{\angstrom}$ is the dipole moment of water, $\lBz$ is the already mentioned Bjerrum length in vacuum and $\sigma_{\LJeff} \simeq \SI{2,5}{\angstrom}$ is the effective distance between the centers of an ion and a water molecule bound to it. The typical dipole-ion interaction energy is much larger, in absolute value, than the typical dipole-dipole interaction energy, which amounts to $-\frac{\delta^2 \lB}{\sigma_{\LJeff}^3} \simeq -5 $.
Adding the Lennard-Jones repulsion, the energy gain per water molecule amounts to $64\,\kB T$, so that the formation energy $u_n$ of an $n$-mer can be eventually estimated to $-64\,n\,\kB T$ (turquoise line in Figure \ref{nmers}c).

In order to improve this energy estimate, we take into account also dipole-dipole electrostatic and steric interactions. To do so, we consider an ion and $n$ water molecules,
fix the ion position at the origin and look for the minimum-energy configuration. We forbid water molecules to assume negative $z$ coordinates in space, i.e. to go beyond the wall. Minimizing through simulated annealing confirms that water prefers to stay close to the ion, with dipoles oriented radially and directed outwards. The so formed $n$-mers, for $n = 4$, 5 or 6 are depicted in Figure \ref{nmers}c, together with the single-$n$-mer energies (dark cyan squares). In the following, we refer to water molecules whose dipole moments are parallel to the plane as ``coplanar'' molecules, and to the water molecule lying on top of the ion, with dipole moment perpendicular to the wall, as the ``top'' molecule.  

Assuming these minimum-energy structures to be the lattice units of an infinite 2D crystal of $n$-mers, one can eventually compute the energy per $n$-mer associated to the formation of such crystal, starting from a crystal of ions only. The procedure is non trivial, mainly because of the long range of Coulombic forces: interactions among dipoles and ions were summed discretely up to a sufficiently large distance, from which a continuous approximation was used. The resulting energies $u_n$ (green circles in Figure \ref{nmers}c) allow to derive the energy gain in adsorbing a water molecule onto an $n$-mer to form an ($n$+1)-mer. Notice that $u_n/n$, i.e.\@ the formation energy of an $n$-mer per unit dipole, is, in absolute value, more than one order of magnitude higher than thermal energy.

\subsubsection{Free energy and $n$-mer distribution}
\label{subsec:freeenergy}
To confirm the locked-water picture proposed in the previous section, we need to compute the free energy gain $\Delta F_{n \rightarrow n+1}$ in adsorbing a water molecule from the bulk on an $n$-mer, to form an ($n$+1)-mer. We will then check that at strong confinement it is always more favorable for a water molecule to be adsorbed on an ion than to stay in the bulk. The mentioned free energy gain can be calculated as
\begin{equation}
\label{eq:DeltaF}
\Delta F_{n \rightarrow n+1} = U_{n+1} - U_n - T(S_{n+1} - S_{n}) - \mu ,
\end{equation} 
where $U_n$ is the average potential energy of water molecules in an $n$-mer, $S_n$ is their entropy, and $\mu$ is the chemical potential of water, i.e. the increase in free energy when a water molecule is moved from the reservoir to the system. What we call here \textit{free energy}, by an abuse of terminology, is properly speaking the \textit{grand-potential}, i.e.\@ the thermodynamic potential associated with the grand-canonical ensemble, by which we describe water. Since the situation is grand canonical for water, and canonical for ions (fixed by electro-neutrality), one sometimes uses the terminology of semi-grand-canonical ensemble.

 The average energy of an $n$-mer is $U_n = u_n + \frac{5}{2}\,n\,\kB T$:  $u_n$ is for the minimum-energy configuration (green circles in Figure \ref{nmers}c, at our best estimate), plus a contribution per water molecule of $\frac{1}{2} \kB  T$ for each degree of freedom (3 in real space and 2 in the dipole moment space), assuming that the Hamiltonian can be expanded quadratically around its minimum.

\begin{figure*}[t]
\centering
\includegraphics[width=0.32\textwidth]{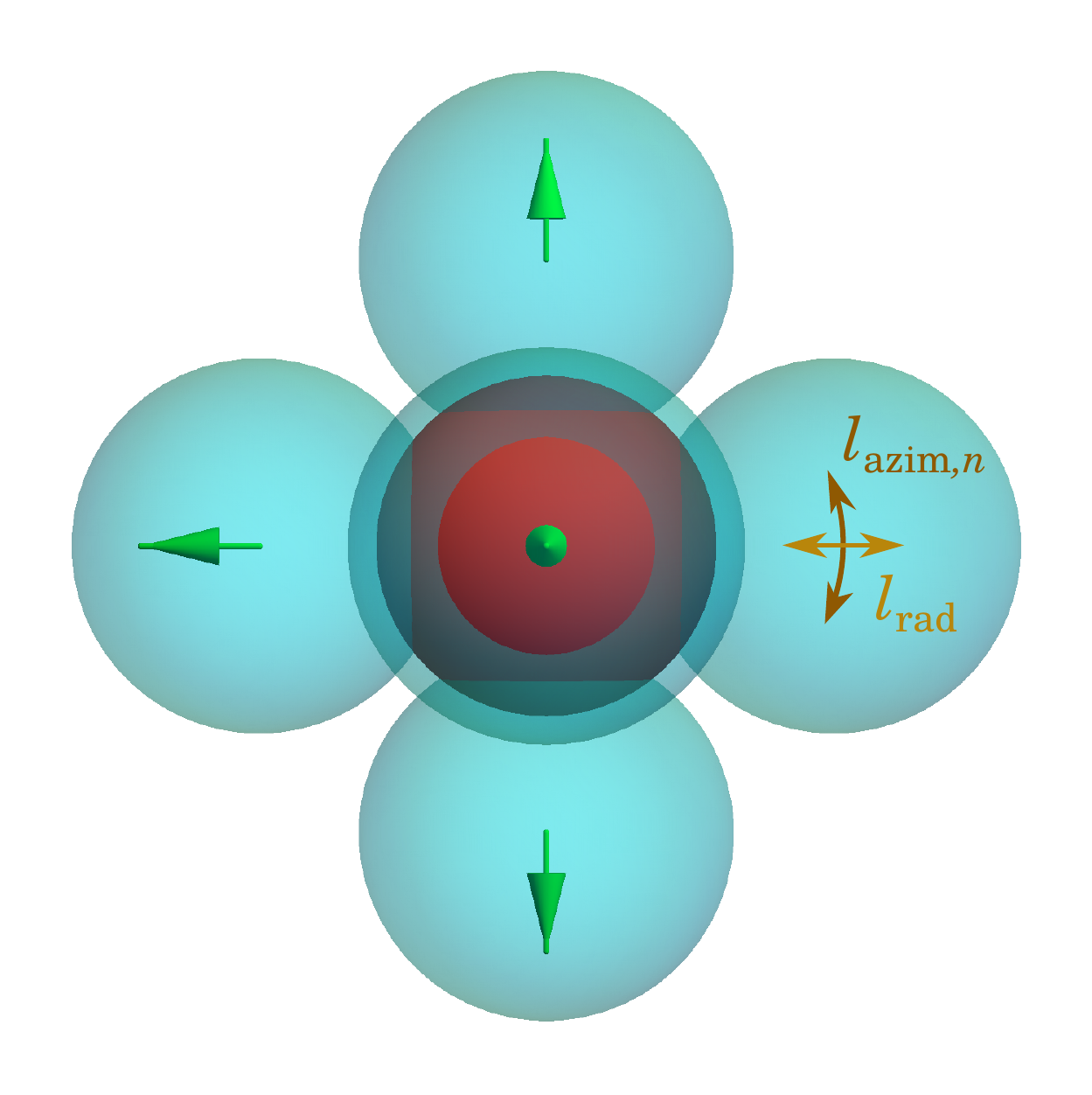}
\includegraphics[width=0.32\textwidth]{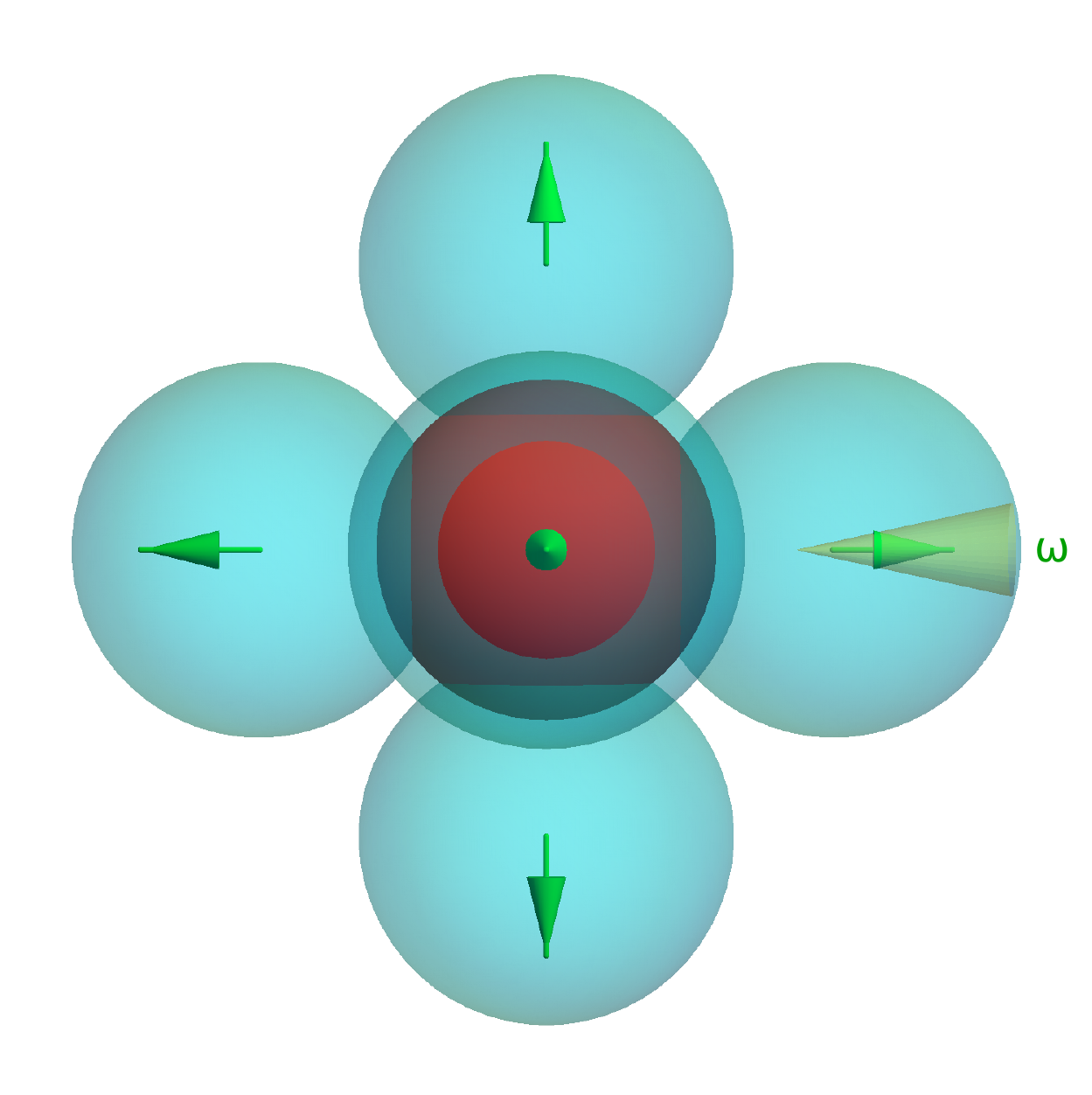}
\caption{Top view of two 5-mers (the charged wall lies below the page). On the left, we show the typical fluctuation lengths $l_{\mathrm{rad}}$ in the radial direction (closer or further from the ion) and $l_{\mathrm{azim},n}$ in the azimuthal direction (around the ion). On the right, we show the typical solid angle $\omega$ spanned by a dipole due to thermal fluctuations. The length of the two arrows and the size of the cone are not to scale.}
\label{fig:5merConfinement}
\end{figure*}
 
 The entropic term $S_n$ can be estimated by considering that a water molecule on an $n$-mer is confined within a volume $v_n = l_{\mathrm{rad}}~ l_{\mathrm{azim,n}}^2 $ and that its dipole moment is also confined within a solid angle $\omega$ (see Fig. \ref{fig:5merConfinement}). $l_{\mathrm{rad}}$ is the small radial distance a water molecule can travel further or closer to the ion with a variation in potential energy of the order of $\kB  T$; it can be estimated within a single water molecule approximation as $l_{\mathrm{rad}} \simeq \sqrt{\pi}~[-\frac{1}{2} U''_{\LJcd}(\sigma_{\LJeff}) + 3 q \lB \delta \sigma_\LJeff ^{-4}]^{-1/2} \simeq \SI{0.17}{\angstrom}$, where  $U_{\LJcd}$ is the charge-dipole (ion-water) Lennard-Jones interaction. In turn, $l_{\mathrm{azim},n}$ is the distance measuring fluctuations of a water molecule along the azimuthal direction around the ion. It can be estimated by fixing positions and dipole moments of all water molecules within the $n$-mer as in the minimum-energy configuration, except for one probe molecule, which is let free to move on the sphere at constant distance $\sigma_\LJeff$ from the central ion. If our probe is a coplanar molecule we have $l_{\mathrm{azim},4}=\SI{3.1}{\angstrom}$, $l_{\mathrm{azim},5}=\SI{1.0}{\angstrom}$ and $l_{\mathrm{azim},6}=\SI{0.30}{\angstrom}$ (of course, azimuthal confinement grows with $n$). Lastly, fluctuations in the orientation of the dipole can be estimated by considering that a deviation of the dipole moment of an angle $\delta\theta$ from the equilibrium position produces an increase $\frac{q \lB \delta}{\sigma_\LJeff^2} (1 - \cos(\delta\theta)) $  in the dimensionless charge-dipole energy, so the solid angle $\omega$ corresponding to an energy increase $\kB  T$ is $\omega= \frac{2 \pi \sigma_\LJeff^2}{q \lB \delta} \simeq 0.087$. 
 
Eventually, estimating entropy $S_n$ as $\kB  \ln \left[ \left( \frac{v_n}{\Lambda^3}\right)^n \omega \right]$, where $\Lambda$ is the De Broglie thermal wavelength used in $\mu$, Eq.~\eqref{eq:DeltaF} can be finally rewritten as
\begin{equation}
\Delta F_{n \rightarrow n+1} = u_{n+1} - u_n + \frac{5}{2}\kB  T - \kB  T  \ln\left( \frac{v_{n+1}^{n+1}}{\Lambda^3 v_n^n} \omega\right) - \mu\,,
\end{equation}
which is independent from $\Lambda$, as it should. While $\beta\Delta F_{3\rightarrow4}$ and $\beta\Delta F_{4\rightarrow5}$ are several tens below zero ($<-30$ in our estimates), $\beta\Delta F_{5\rightarrow6}$ nearly vanishes ($-1.1$ in our estimate). This suggests that available water molecules must be adsorbed on ions until every ion has 5 water molecules; beyond that point, it becomes in practice equally favourable for water molecules to be adsorbed on a 5-mer and form a 6-mer or to remain in the bulk. This is indeed shown by an analysis of the composition of $n$-mers as a function of distance $D$ (Figure \ref{nmers}b).

\subsubsection{Thermal fluctuations of bound water}
\label{sec:ptheta}

An suggested by the relatively high energies involved in the process, the $n$-mer formation significantly impacts the orientational mobility of water, that is strongly localized close to the ion. At finite temperature, however, water is not completely locked in its ground state configuration and acquires some freedom to move around its equilibrium position. Starting from minimum-energy configurations for $n$-mers that include interactions with the wall, we analyze the importance of these fluctuations, by computing the probability distribution of the orientation angle of dipoles within the same $n$-mer. We are thus scrutinizing a fine property, reflecting the effect of temperature on mutual interactions among ions, water molecules and wall altogether. 

\begin{figure*}
	\centering
	\includegraphics[width=0.95\textwidth]{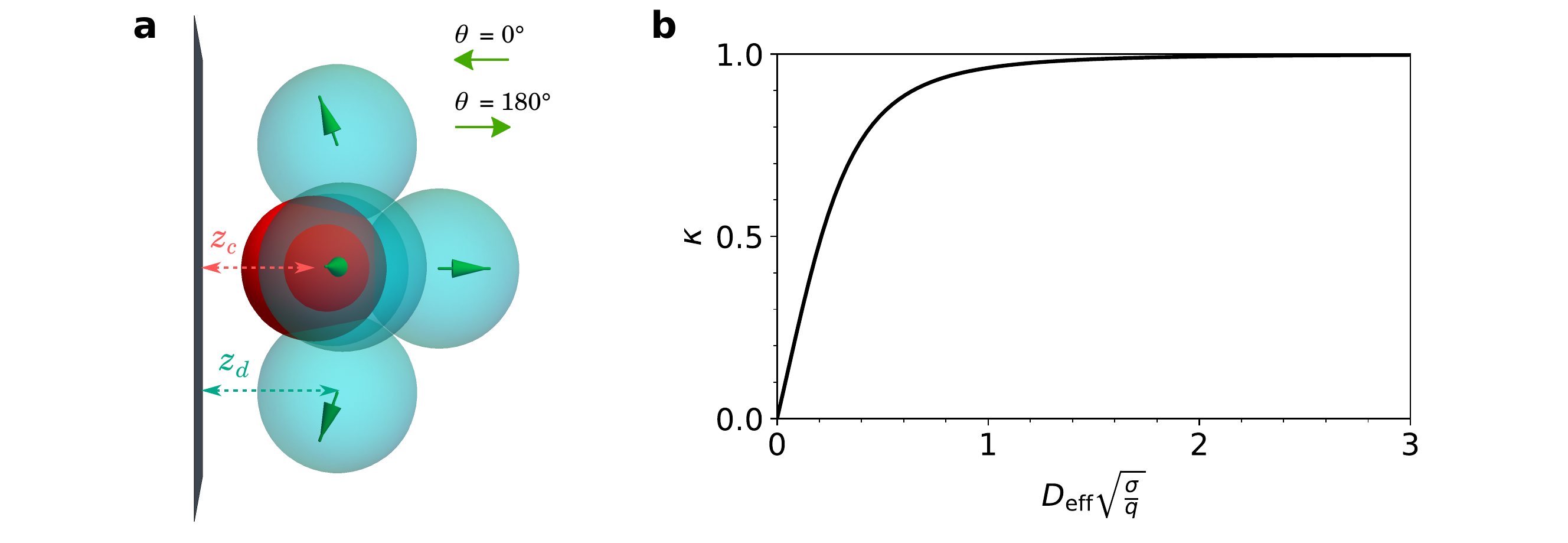}
	\caption{Minimal energy configuration of an $n$-mer and the effective field felt by $n$-mers. (a) Side view of the minimum-energy configuration of 5-mer, interacting with the wall through Lennard-Jones and electrostatic potentials. $z_c$ is the distance of the ion from the wall and $z_d$ is the distance of a coplanar water molecule from the wall. (b) Effective field $\kappa$ as a function of the dimensionless distance between ion layers $D_\eff \sqrt{\sigma/q}$, according to Eq.~\eqref{eq:kappa}. This corresponds to the field felt by an ion at contact with the wall, due to the presence of all the surrounding ions. It tends to zero at vanishing distance, when a uniform ion distribution is expected; it tends to unity (i.e.\@ to the single bare wall field) at infinite distance, when correlation with the opposite ion layer is lost.}
	\label{fig:kappa}
\end{figure*}

We fix $n-1$ water molecules to their minimum-energy positions and orientations, and let the remaining one free to move and rotate. Marginalizing numerically with respect to the three spatial degrees of freedom and one of the rotational degrees of freedom, one obtains a probability distribution $p(\theta)$ for the angle $\theta$ formed by the dipole moment with respect to the normal of the plane (see Fig. \ref{fig:kappa}a). This can be done using as probe any of the $n-1$ coplanar molecules or the top molecule, for different $n$. 

\begin{figure}
	\centering
	\includegraphics[scale=.7]{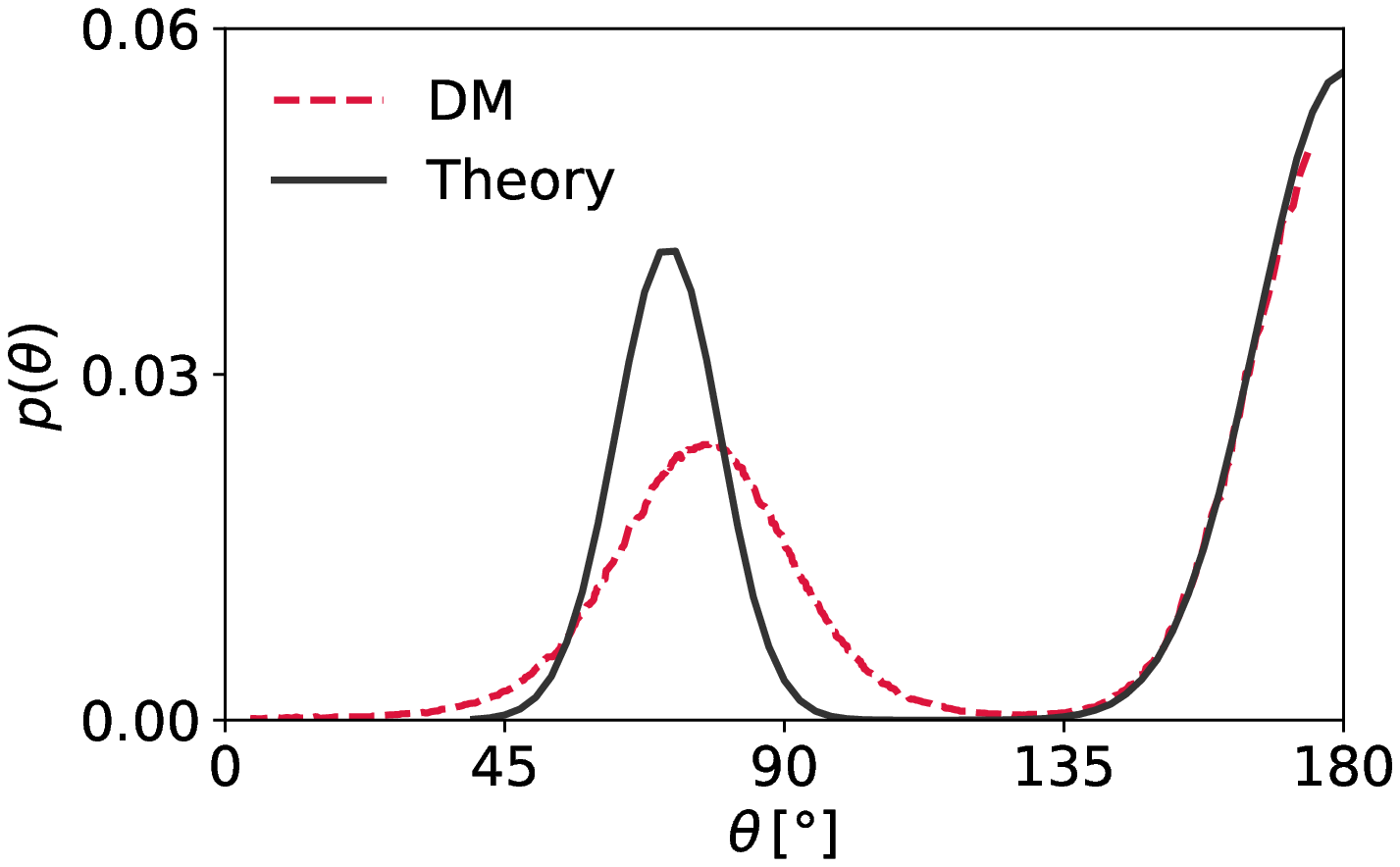}
	\caption{The probability distribution of the angle between the dipole moment of water in an $n$-mer and the surface normal at $D=\SI{40}{\angstrom}$. The angle $\theta$ is with respect to the nearest surface: $\theta=\SI{0}{\degree}$ pointing at the surface on which the $n$-mer resides and $\theta=\SI{180}{\degree}$ pointing directly away (see Fig. \ref{fig:kappa}a). Contrary to bulk liquid water, the dipole moment is not free and is limited to specific orientations. Due to non-zero temperature, the dipoles fluctuate around these minimum energy positions. The peak at $\theta\simeq\SI{80}{\degree}$ represents coplanar water molecules, and the one at $\theta=\SI{180}{\degree}$ top water molecules (see Fig. \ref{fig:kappa}a). At this distance, the former are $\sim4.5$ times as many as the former, which would be evidenced by a plot of $p(\theta)\sin\theta$.}
	\label{fig:orientation}
\end{figure}

In Fig. \ref{fig:orientation} we compare the result of this analysis with simulations using only the dipole interactions for water (as discussed in Methods) for $D=\SI{40}{\angstrom}$. Indeed, the minimum-energy $n$-mer configuration is computed at $\kappa=1$ (i.e.\@ $D_\eff\gg\sqrt{q/\sigma}$, see Fig. \ref{fig:kappa}b), but results seem to be robust with decreasing $D$. The theoretical distribution shown is a weighted sum of the distributions computed for coplanar and top dipoles, at $n=5$ and 6, using the fact that the fraction of 5-mer and 6-mers is known (Figure \ref{nmers}b). Since this calculation does not allow for cooperative fluctuations away from the ground state, it predicts a distribution that is more sharply peaked, but the result still matches the simulation closely.

\subsubsection{Limited resources argument}
\label{sec:limitedresources}

\begin{figure}
    \centering
    \includegraphics[scale=0.7]{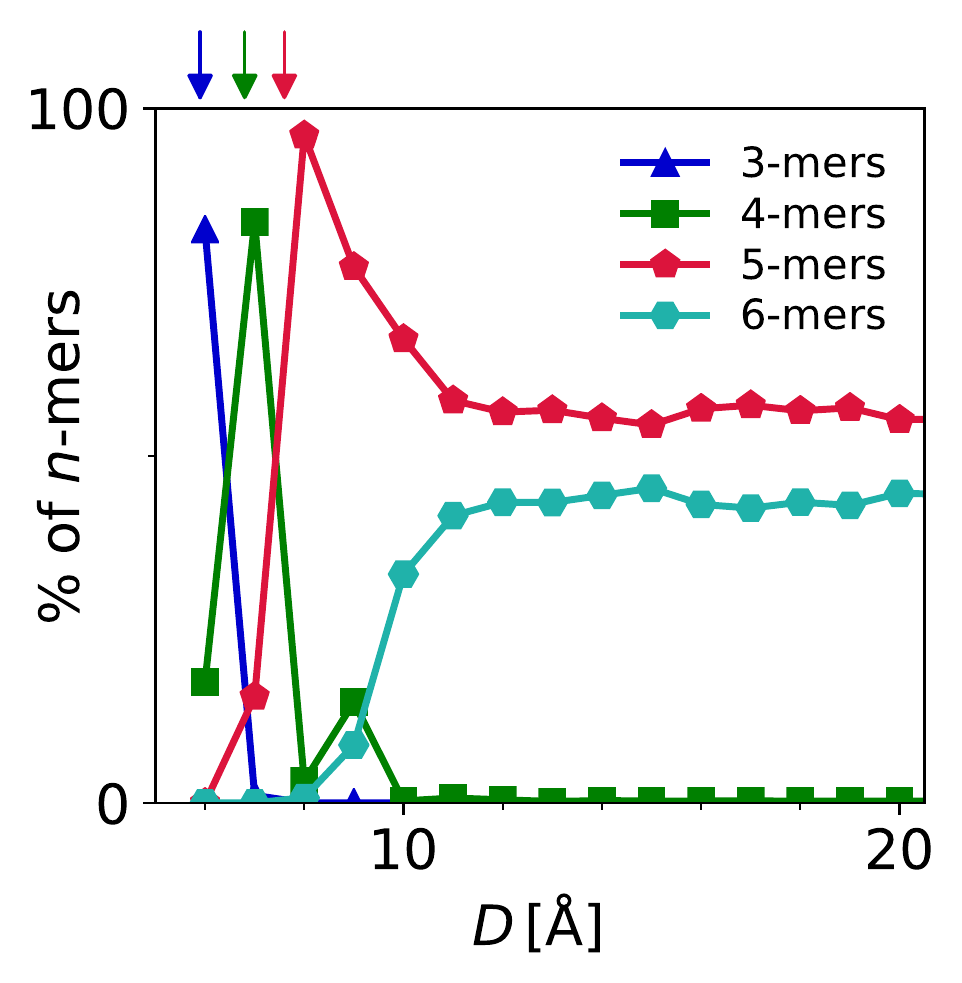}
    \caption{Equivalent of Figure \ref{nmers}b for simulations with only dipolar water interactions, showing for distances $D$, percentage of 3-, 4-, 5- and 6-mers observed. Arrows show analytical predictions for $D_n$, given by Eq.~\eqref{eq:FractionBoundWater}: these correspond to the peaks of the curves shown.}
    \label{nmersDistributionDM}
\end{figure}

Supposing all water molecules to bind to Ca$^{2+}$ ions at small $D$, one can define precise distances $D_n$ at which all ions are bound to exactly $n$ water molecules. At such distances, the following limited-resources equation holds, expressing the fact that the number of water molecules in the pore per unit surface (l.h.s) should equal $n$ times the number of ions per unit surface on both sides (r.h.s.):
\begin{equation}
\label{eq:limitedresources}
\rho_{\mathrm{w}}\,D_n = 2\,n\,\frac{\sigma}{q} \,.
\end{equation}
Here, $\rho_\mathrm{w}$ is the total number of water molecules divided by the total simulation volume ($D$ times the surface), so that $\rho_{\mathrm{w}}\,D$ is the surface density of water. $\rho_\mathrm{w}$ is a measurable function of $D$, but it can also be estimated supposing that in the volume effectively available to water (the region at least a Lennard-Jones unit far from walls and ions) the water density is constant (this is true within a 17\% error for the considered $D$ range). Anyway, using Eq.~\eqref{eq:limitedresources} and measured values of $\rho_\mathrm{w}$, one obtains $D_3=\SI{5.9}{\angstrom}$, $D_4=\SI{6.8}{\angstrom}$ and $D_5=\SI{7.6}{\angstrom}$. These values are in quantitative agreement with peaks in the observed number of 3-mers, 4-mers and 5-mers as a function of distance (Figure \ref{nmersDistributionDM}). 

An interesting observable to look at is the fraction of water molecules bound to ions $f(D)$, shown in Figure \ref{nmers}e. In light of our present discussion, we can provide an analytical description of such curve:
\begin{equation}
\label{eq:FractionBoundWater}
f(D)= \begin{dcases}
\ 1 & \text{if}\  D\leq D_5\\
\ \frac{10}{10+\frac{q}{\sigma}\rho_\mathrm{b}(D-D_5)} & \text{if}\  D>D_5
\end{dcases}
\,,
\end{equation}
where $\rho_\mathrm{b}$ is the density of bulk water. The factor 10 emerges from the fact that, in a surface $q/\sigma$ hosting one ion per wall, no more than 10 water molecules (5 on each ion) can be bound. This approximation discards the differences between 5-mers and 6-mers, a valid approximation for our purposes. Also neglected is the presence of a few water molecules, bound to the walls. Nonetheless, this limited resources argument seems to capture all the  physical ingredients relevant to explain the numerical curve.

\subsection{Equation of state}
\label{sec:eos}

\subsubsection{Ion densities}
\label{sec:eosDensities}

In order to compute the density profile and eventually the equation of state, we refined the $n$-mer minimum-energy configurations discussed in \ref{subsec:energy} by introducing Lennard-Jones and electrostatic interactions with the wall, for the ion and the $n$ water molecules. A numerical minimization of the same kind as the one in SM section \ref{subsec:energy} allows to identify the distances from the wall $z_{c}$ and $z_{d}$ at which, respectively, ion (\textit{c}harge) and coplanar water molecules (\textit{d}ipole) prefer to stay. These are better defined by Fig. \ref{fig:kappa}a. The shape of $n$-mers is qualitatively unchanged, were it not for the fact that 1) ions, charged and slightly smaller, penetrate closer to the wall than water molecules and 2) the dipole moments of coplanar molecules are now slightly tilted toward the wall ($\theta<\SI{90}{\degree}$ in the Figure). For $n=5$, $z_{c}=\SI{2.25}{\angstrom}$ and $z_{d}=\SI{2.72}{\angstrom}$ and these quantities vary of only a few percent with $n$. The importance of $z_c$ lies in that it defines the effective distance $D_\eff=D-2z_c$ between the two planes where ions are positioned.

In order to retrieve ion densities, we treat $n$-mers in their just described minimum-energy configuration as effective charged objects and we use a modified version of the correlation-hole theory described in \cite{Samaj2018}, where we introduce a soft potential $U_\LJwall$ between charges and wall. This potential is the sum of the ion-wall Lennard-Jones potential and of $n-1$ water-wall Lennard-Jones potentials.

The theory \cite{Samaj2018} is based on the fact that the effective electric field $\kappa / (\beta q e \muGCz)$ felt by ions lying on one wall is due to the presence of a staggered equal arrangement of ions on the opposite wall (from this perspective, the fields exerted by the two bare walls cancel out exactly). Here, $\muGCz = (2\pi q \lBz \sigma)^{-1}$ is the Gouy-Chapman length in vacuum, while $\kappa$, the dimensionless effective field, is a  monotonic function of the distance between the two planes where ions lie. 
It is useful to recall that $\kappa$ must be 1 at infinite distance, when ions lying, say, on the left wall feel the presence of the left wall only: since the right wall and its counterions are indeed infinitely far, inter-layer correlation disappears and the right and left half-systems are electroneutral and do not interact. Also, $\kappa$ must go to 0 at distances $D_\eff\ll\sqrt{q/\sigma} \simeq \SI{8}{\angstrom}$, when ions tend to the uniform distribution along $z$ and are strongly correlated along $xy$ to form a single Wigner crystal \cite{GoPe96,Varenna}. For intermediate distances, the function $\kappa(D_\eff)$ we use is based on a correlation-hole approach (so-called ch2 in \cite{Samaj2018}), that has been shown to yield very good results for the liquid ($\Xi\lesssim31\,000$) and the crystal phase ($\Xi\gtrsim31\,000$) in the case of point-like ions interacting with hard walls. Within this approximation, the dimensionless effective field, plotted in Figure \ref{fig:kappa}b, is given by
\begin{equation}
\label{eq:kappa}
\kappa(D_\eff)=\frac{D_\eff \sqrt{\frac{\sigma }{q}}}{\sqrt{\frac{\sigma}{q}D_\eff^2+\frac{1}{2 \pi  \left(D_\eff \sqrt{\frac{\sigma }{q}}+1\right)}}}\,.
\end{equation}

Now, if we account for Lennard-Jones interactions with the wall, the density $\rho$ of effective charges as a function of distance $z$ between wall and central ion (which is nothing but the ionic density), is given by
\begin{equation}
\label{eq:rho}
\frac{\rho(z)}{2 \pi \lB \sigma^2}=\mathcal{N(D)}(e^{-\kappa(D_\eff) \frac{z}{\muGCz} - \beta U_\LJwall(z, D))} + 
  e^{-\kappa (D_\eff) \frac{D - z}{\muGCz} - \beta U_\LJwall (D - z, D)}) ,
\end{equation}
where $\mathcal{N(D)}$ is a normalization constant, ensuring electroneutrality. 

Results of Eq.~\eqref{eq:rho} are compared with simulations in Figure~\ref{theoryPredictions}a for $D=\SI{8}{\angstrom}$ and in Figure~\ref{theoryPredictions}b for $D=\SI{12}{\angstrom}$. Notice that this approximation considers $n$-mers to be rigid objects and neglects the fact that a $\kappa\neq 1$ can (slightly) modify the $n$-mer's configuration, namely $z_c$ and $z_d$. For simplicity, top water molecules (the top dipole in an $n$-mer) are not considered in $U_\LJwall$.  Most importantly, this approximation is not valid at large distances (see SM section \ref{sec:DielectricConstant}), where free water fills the pore and screens electrostatic interactions -- in other words, the Bjerrum and Gouy-Chapman lengths are not constant with $D$.

\subsubsection{Pressure}
\label{sec:eosPressure}

Once densities are known, one can compute the pressure using the contact theorem, an exact result relating pressure with the ion density at contact with a hard charged wall \cite{BlHL79,Andelman2010}. We extend this equality in the following way, to account for soft interaction with the wall:
\begin{equation}
\label{eq:pressurecontacttheorem}
P(D)= -\frac{e^2 \sigma^2}{2 \varepsilon_0} + \int_0^\frac{D}{2} F_\LJwall(z) \rho(z) \diff z \,.
\end{equation}
Here, $F_\LJwall(z)$ is the force exerted on both walls by an $n$-mer in $z$ and corresponds to the derivative of $U_\LJwall(z)$. The sign convention is that positive contributions represent repulsion and negative ones attraction.

The first term in Eq.~\eqref{eq:pressurecontacttheorem} represents the minimum pressure attainable due to electrostatic interactions, while the second one, always positive, is due to ions pushing against the walls. The predicted pressure is represented in Fig. \ref{theoryPredictions}c and has a minimum at $D_{\mathrm{min}}$ close to $\SI{6}{\angstrom}$, corresponding to $P=\SI{-6.5}{\giga\pascal}$. For smaller distances, the Lennard-Jones repulsion with the wall starts to play a crucial role: pressure increases and eventually becomes positive. This increase must not be mistaken by the pressure increase observed in the hard-wall point-like-ion situation \cite{Samaj2018}, in which case pressure would continue to drop down to a value twice as negative and then increase much more abruptly upon decreasing further the distance. For distances $D>D_\mathrm{min}$, the curve is given by:
\begin{equation}
\label{eq:pressureanalytical}
\frac{\beta P(D)}{2 \pi \lB \sigma^2}= -1+ \kappa(D_\eff)\left( \frac{1+e^{\kappa(D_\eff) \frac{D_\eff}{\muGCz}} }{ 1-e^{\kappa(D_\eff) \frac{D_\eff}{\muGCz} }}  \right)\, .
\end{equation}    

\begin{figure}
	\centering
	\includegraphics[scale=0.6]{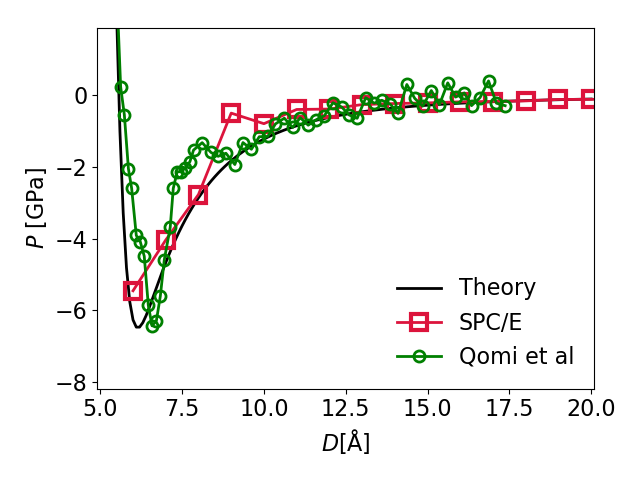}
	\caption{Pressure $P$ as a function of bare distance between walls $D$. The theoretical curve, computed from Eq.~\eqref{eq:pressurecontacttheorem} and obeying  Eq.~\eqref{eq:pressureanalytical} for $D>6\,$\AA\, is compared to SPC/E simulations (as in the main text) and to atomistic simulations of a Tobermorite crystal taken from \cite{Masoumi2017}. Pressure data are given in \cite{Masoumi2017} as a function of the distance between the centers of mass of the two solid crystalline walls, also modelled at atomistic level: in order to be presented on this graph, they were shifted to the left by \SI{6.7}{\angstrom}, which makes the effective distance between ion layers approximately equal to our $D_\eff$.} 
	\label{fig:PressureQomi}
\end{figure}

This model does not account for solvent layering, which is probably responsible for the non-monotonicity of the curves extracted from numerical simulations. These are plotted in Fig. \ref{fig:PressureQomi}, together with the theoretical prediction from Eq.~\eqref{eq:pressurecontacttheorem} and with atomistic simulations of Tobermorite \cite{Masoumi2017}. {Atomistic studies that use a more specific C-S-H model have, typically, a higher surface charge density, which should correspond to a higher strength, but this effect could be limited by the presence of surface heterogeneities and other ion types \cite{Masoumi2019}.} The good agreement with the atomistic simulations of Ref. \cite{Masoumi2017} is noteworthy, especially remembering that the pressure curve reported there was rationalized by a 7-parameter fit. This is at variance with our theoretical approach, that does not involve any fitting parameter.
In Figure \ref{theoryPredictions}c of the main text, the theoretical curve is also compared to primitive model simulations run in vacuum. The latter data obey \textit{a fortiori} Eq.~\eqref{eq:pressureanalytical}, with $D_\eff = D - 2 z_c$ and $z_c\simeq\SI{2.2}{\angstrom}$, as given by a simple balance between electrostatic attraction to the wall and Lennard-Jones repulsion from it. 

Using the saddle point method to estimate $\mathcal{N(D)}$ and the integral in Eq.~\eqref{eq:pressurecontacttheorem}, it is possible to write the pressure as
\begin{equation}
\beta P(D)= -2\pi\lBz\sigma^2 + \frac{\sigma}{q}\beta F_\mathrm{LJ,w}(z_0)\,,
\end{equation}
where $z_0$ is the extremum point of the function appearing to exponential in the Boltmzann factor, that is $\kappa \frac{z}{\muGCz}+\beta U_\LJwall(z)$. Since $\kappa$ and $F_\LJwall$ depend on $D$, $z_0$ does too. The interpretation of this formula is straightforward: since $n$-mers are concentrated at a distance $\simeq z_0$ from the closest wall ($z_0 = z_c$ at large $D$), the pressure they exert on the walls is the force $F_\LJwall$ divided by the surface $\frac{q}{\sigma}$ pertaining to each of them. Agreement with the curve calculated from \eqref{eq:pressurecontacttheorem}, in black in Fig. \ref{fig:PressureQomi}, is perfect.

\section{Water properties}
\subsection{Water model}
\label{sec:waterModel}
{
Despite significant advances in the past century, water remains a challenging material to model, as evidenced by the plethora of different models developed during that time. In this study, we have opted to use a relatively simple representation of water: the SPC/E model \cite{Berendsen1987}. This model treats water as a rigid molecule, with 3 partial charges and one Lennard-Jones site, which is the minimum degree of complexity needed for hydrogen bonding and a tetrahedral structuring. Nonetheless, it is known that the SPC/E water model effective at capturing the structure and dynamics of bulk water \cite{Mark2001}. 

One weakness of the SPC/E model is that starts to deviate from experimental results in confined or high pressure situations, where the TIP4P/2005 model performs better \cite{Abascal2005,Dix2018}. Having understood that the phenomena of interest here are related to water-ion structuring and are prevalent when ion-water correlations become dominant on water-water correlations, significant changes in our picture would require dramatic differences in the water model, well beyond the range of the most used ones. Hence we do not expect that our results are qualitatively changed by the more accurate water description of TIP4P/2005. To verify this, we performed additional simulations using the TIP4P/2005 water model, for $\sigma=3e/\si{\nano\meter}^2$ and separations $D=6$, $8$, and $12 \si{\angstrom}$ and the same procedure (with due differences in terms of longer simulation times required) for preparation and equilibration of the samples. These simulations showed that the structuring of the ions and the resultant net cohesion between the C-S-H surfaces with the TIP4P/2005 model was very similar to what was obtained with the SPC/E model (Fig.~\ref{fig:tip4p}).

\begin{figure*}
	\centering
	\includegraphics[width=\linewidth]{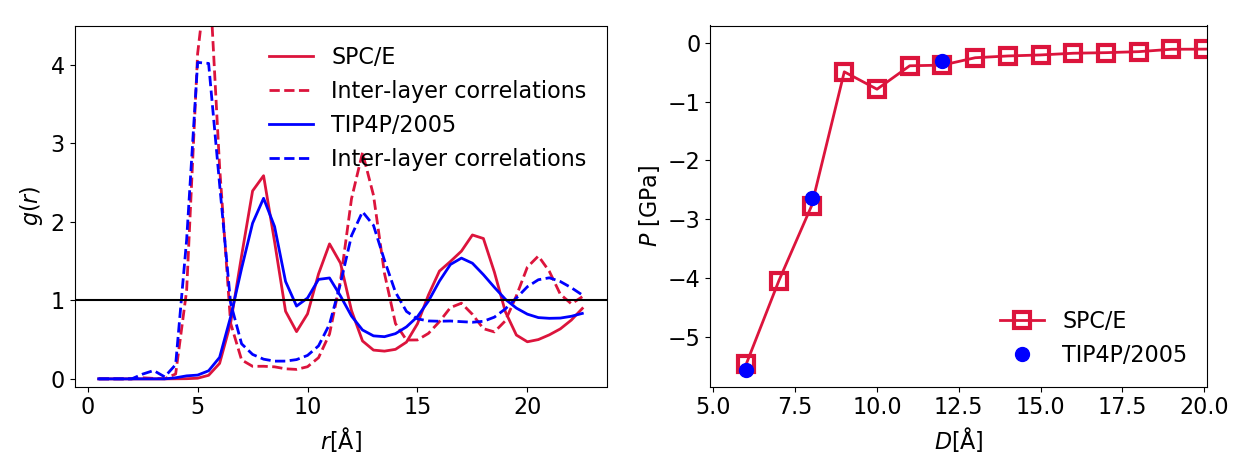}
	\caption{ Comparison between results with SPC/E and TIP4P/2005 models for water at $\sigma=3e/\si{\nano\meter}^2$. The pair correlation $g(r)$ calculated at $D=8\,\si{\angstrom}$ shows that differences in the water model only slightly alter ion structuring, and the net pressure between the confining surfaces is approximately the same. }
	\label{fig:tip4p}
\end{figure*}

A final consideration is the effect of polarizability in the water molecule. While one might question whether the interlocked and nearly solid ion-water structure we observe would disappear once polarizability is accounted for, recent studies with polarizable water models for clays have, in fact, demonstrated that the polarizability enhances the slowing down of the water molecule dynamics, indicating that these effects would rather work to confirm our picture \cite{LeCrom2020}. Another study comparing common polarizable and non-polarizable water models confined between MgO surfaces showed that interfacial water structure and orientational patterns were similar, but the polarizable models exhibit more constrained degrees of freedom and longer-ranged water layering---again indicating that polarizability would enhance the effects discussed in our work \cite{Kamath2013}. To summarize, in the interests of developing a coarse-grained, semi-atomistic approach, the SPC/E water model is overall a reasonable choice.
}
\subsection{Dielectric constant}
\label{sec:DielectricConstant}
The dielectric response of a material can be quite complex at the atomic level. While the relative dielectric constant $\varepsilon_r$ is a macroscopic quantity, it arises from this complex microscopic behavior. Even ignoring how the confinement in our system would change the macroscopic $\varepsilon_r$, trying to use this $\varepsilon_r$ for interactions at the nanoscale (as in the PM) has a host of problems. With numerous charges enclosed in a small volume, the polarization of the solvent would depend non-trivially on the arrangement of all the ions and solvent molecules, so taking it to behave the same as in the macroscopic material exposed to an external field is a very strong assumption.

Explicit inclusion of the solvent allows one to directly incorporate this as a microscopic phenomenon. Unfortunately, this is a feature that is actually quite difficult to capture correctly, and many water models that give otherwise similar results produce drastically different values for the dielectric constant \cite{Sprik1991}. However, while it is difficult to be confident in a precise value, general trends can be informative. In MD simulations, a standard way to compute $\varepsilon_r$ is from the total dipole moment, $M$ \cite{Gray1986}. $\varepsilon_r$ can be related to $M$ through the fluctuation-dissipation theorem. Specifically, $\varepsilon_r = 1 + \chi$, where $\chi$ is the electric susceptibility, and the fluctuations in $M$ are related to its dissipation through $\chi$. For an isotropic system, $\langle M \rangle=0$ and the variance of $M$ is simply $\langle M^2\rangle$,  giving
\begin{equation}
\chi = \dfrac{1}{\varepsilon_0 V\kB T} \dfrac{\langle M_x^2+M_y^2+M_z^2 \rangle}{3} \,.
\end{equation}

\begin{figure*}
	\centering
	\includegraphics[width=.45\linewidth]{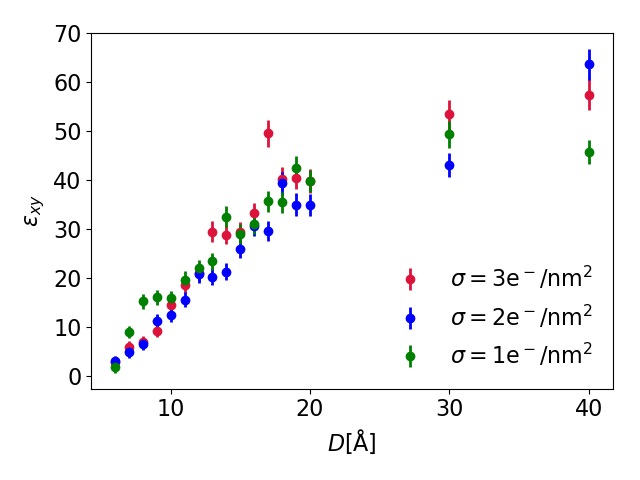}
	\includegraphics[width=.45\linewidth]{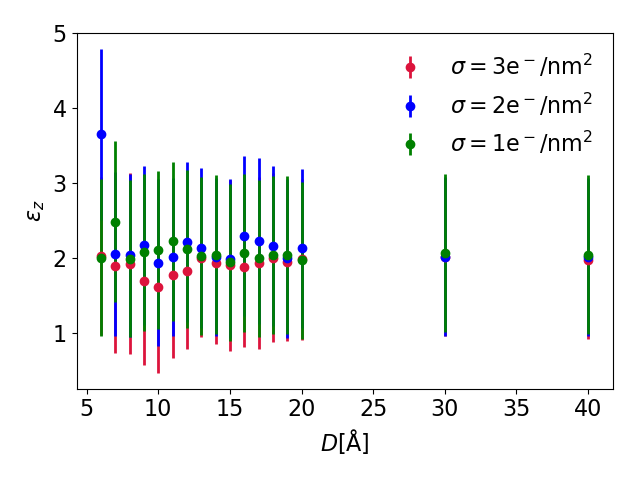}
	\caption{The transverse (left) and perpendicular (right) relative dielectric constant of water in our simulations, as a function of $D$. This is computed from the fluctuations of the total water dipole moment. Due to the large coupling between ions and water at small separation, the effective dielectric screening is far lower than the bulk value. The dielectric tensor remains anisotropic even for largest separations considered, which is consistent with experimental measurements of a slow decay in dielectric properties with distance \cite{Fumagalli2018}. }
	\label{fig:dielectric}
\end{figure*}

The behavior of the dielectric constant under confinement is not fully understood. The anisotropy introduced by the slab geometry leads to differing behaviors for $\varepsilon_{xy}$ and $\varepsilon_z$. $\varepsilon_{xy}$ can be computed in the same way as the bulk calculation except only considering the $x$ and $y$ components of $M$, while $\varepsilon_z$ needs to be reformulated for the very different boundary conditions \cite{Froltsov2007}:

\begin{equation}
\varepsilon_{xy} = 1+\dfrac{1}{\varepsilon_0V\kB T} \dfrac{\left\langle M_x^2+M_y^2 \right\rangle}{2} 
\end{equation}
{
\begin{equation}
\varepsilon_z^{-1} = 1- \dfrac{\left\langle M_z^2 \right\rangle}{\varepsilon_0V\kB T}  \,.
\end{equation}
}
Using this, we calculate an effective dielectric constant from our simulations (Fig. \ref{fig:dielectric}). Though the precise values obtained depend on the model for water, the relative decrease of dielectric constant in confinement is revealing. When all the water is bound to ions, it is highly localized and unable to reorient, leading to a drastically lower dielectric constant than at larger separations \cite{Schlaich2019}. This lends credence to the locked water picture in which the water is unable to screen electrostatic interactions and helps explain the large increase in net attraction.

{
The calculation of a dielectric constant is useful to understand, at a qualitative level, how the screening is altered by confinement. However, it is important to note that the effects of water structure and dynamics at a microscopic level cannot be fully captured in a single number in many cases. {While it has been proposed that, with $\Xi$ of order a few tens, the results of explicit solvent simulations could be matched by rescaling the bulk dielectric constant in the PM \cite{Pegado2008a}, here we find a very different picture, at higher surface charges and for more confined systems, because strong water structuring effects arise.} This is seen in the $D$-dependent, anisotropic dielectric properties and increased cohesion we observe, as well as the hydration-related pressure oscillations (see the pressure spike at $D=8\,\rm \AA$ and $\sigma=1e/\si{\nano\meter}^2$) which have also been reported in a wide array of studies \cite{Claesson1986,Schneck2012,Chen2016}. What the locked water picture shows is that, in the appropriate limit, a new ground state gains relevance, and water becomes so structured that its effect on the cohesion is minimal---an effect obtained not by partially rescaling the dielectric constant but by assuming that water is not free to screen electrostatic interactions at all.
}

\section{A bit of history}


\subsubsection{\textit{De caementorum natura}, or the invention of C--A--S--H}

The invention of modern cement cannot prescind from the discovery of natural cements. These are naturally available sands or mixtures whose mortars feature stronger adhesion than regular lime-based mortars, and, most importantly, hydraulicity: the possibility to set in (sea)-water. While lime mortars seem to have been used by many civilizations millennia BCE in Mesopotamia, Egypt, China and Greece, the first examples of hydraulic natural concrete (sometimes called hydraulic lime) date to 700 BCE. The Nabateans, a bedouin population living between present-day Syria and Jordan, used it to build underground water-proof tanks; their extensive water system (reservoirs, cisterns, aqueduct) allowed them to survive and found settlements in the desert.

The most intense and, at the same time, documented use of natural cement before the modern era is with no doubts to be ascribed to Ancient Romans. Archaeological findings and subsequent scientific analyses \cite{Jackson2013,Jackson2017,Oleson2004} clearly show that roman harbours had docks and submarine breakwaters built in natural cement. Scientific research has focused on Roman harbours of the west coast of the Italian peninsula, from present-day Southern Tuscany down to the Naples area, but the one of Caesarea, now in Israel, also features huge perfectly preserved concrete blocks that have resisted underwater for two millennia. Archaeologists have conjectured that the naval power Romans had conquered over the Mediterranean by the 2nd century CE was due to a large extent to the fact that their harbours were not only cleverly built, but also built with concrete.

Hydraulicity was not the only property of cement Romans were interested in. They had discovered that concrete was also much stronger than common mortars and started using it for public architecture. The most famous example is undoubtedly the dome of Rome's Pantheon, built by emperor Hadrian in the fist quarter of the 2nd century CE. With its \SI{5}{\tonne} and \SI{43}{\meter} of diameter, it is still the biggest unreinforced concrete dome in the world \cite{Moore2010}.   

What were Romans using to make their concrete? In Roman architect Vitruvius' \textit{De Architectura}, written in the second half of the 1st century BCE, a whole book is devoted to building materials. After describing lime and the proper way to make a mortar out of it,
Vitruvius talks about a ``powder'', that under water suddenly absorbs liquid and quickly hardens, emphasizing heat release.
The substance Romans were using, called by Seneca \textit{Puteolanus pulvis} in \textit{Quaestiones Naturales}, goes nowadays by the name \textit{pozzolan}, in Italian \textit{pozzolana}, from the name of the town where it was quarried (Pozzuoli, ancient  \textit{Puteoli}, in the Naples region). It is a natural ash of volcanic origin: it works in a very similar manner as modern cement, in that it undergoes a hydration reaction producing C--A--S--H (Calcium Aluminum Silicate Hydrates), a variety of C--S--H where some silicon has been substituted by aluminum.

It is interesting to note that the word cement, and its translation in most European and many non-European modern languages, derives from Latin \textit{caementum} (in turn from \textit{caedere}, to cut), referring to rubble and smashed stone, mostly tuff, that had to be mixed to pozzolana and calcium hydroxide to form concrete.

Romans' astonishment for the fact that ``dust, the most insignificant part of the Earth'', could ``become a single stone, impregnable to the waves, the moment of its immersion, and increase in hardness from day to day'' (Pliny the Elder, \textit{Naturalis Historia}) did not lead them to understanding much more about its nature. They only knew, and this was enough for any practical purpose, that it was of volcanic origin and that it must have something to do with high temperature environments (`fire'). This is maybe why the description Vitruvius, in \textit{De Architectura}, makes of the exothermic hydration reactions (echoed by Saint Augustine four centuries later in \textit{De Civitate Dei}) appears so amusingly and surprisingly accurate: ``the urgent need of moisture suddenly satiated by water seethes with the latent heat [\textit{calor latens}] in these substances and causes them to gather into a unified mass and gain solidity quickly.''

\subsubsection{The modern era}

After the fall of the Western Roman Empire, cement went back to being practically unknown to architects and builders. To meet cement again in (documented) History, we need to fast-forward to the beginning of the 15th century. Louis XII is king of France and needs to build a bridge over the river Seine in Paris. He asks Venetian architect and clergyman Giovanni Giocondo to develop the project. A man of letters, Giocondo is probably the best living expert of Vitruvius' texts and decides to make use of Neapolitan pozzolan. According to some sources, the then Pont Notre Dasme, inaugurated in 1515 and then completely destroyed and rebuilt through the centuries, presented foundations in natural cement: for the first time after a millennium, pozzolan was being used again for a large-scale work.

Pozzolan is not the only natural earth that produces hydraulic mortars. We know that Dutch builders, at the beginning of the modern era, were using a powder coming from the Eiffel region, between present-day Germany and Belgium. They called it \textit{trass}. Trass (also referred to as terras) had no fortune in commercial exchanges. It would be interesting to understand why, but we will limit ourselves to noticing, as in \cite{Courland2011}, that its name lacked any appeal: to British it sounded too much like \textit{trash}, and to French it resembled the word \textit{travers}, a flaw.

In the 18th century, people finally started to look for a scientific explanation for the fact that, at first, lime could transform from powder to solid rock, upon addition and consequent evaporation of water. The first attempts were not much closer to reality than the Romans'. Réaumur, Macquer, Becher speak either of some sort of gravity effect or of fire hidden in the stones. Among these fuzzy chemical theories, the documented tips on how to make better mortars flourished: engineer Giovanni Antonio Borgnis suggested to dilute quicklime in wine and add lard or fig juice, while some of his colleagues recommended rather ox blood and urine \cite{Simonnet2005}. In 1783 Antoine-Laurent de Lavoisier finally brought order to this babel of outlandish theories, by discovering oxygen and hydrogen and giving water its chemical formula. This was the birth of modern Chemistry.

If somebody has to be considered the inventor of cement, it should be 
British engineer John Smeaton (1724-1792) and his baker. The first user of the word ``horsepower'', before James Watt, and of the expression ``civil engineer'', as opposed to ``military engineer'', Smeaton was the designer of a series of bridges, harbours and canals, and, last but not least, of the famous lighthouse of Eddystone. There existed, and still exists, a dangerous stack of rocks, the Eddystone rocks, south of Plymouth, in the English Channel. Throughout history, many ships had sunk there during tempests and since the end of the 17th century people had tried to build  a lighthouse to warn sailors of their presence. The first two attempts were taken down due to the strength of storms and to fires. When Smeaton was asked to build the third lighthouse, he started experimenting new kinds of mortars that could resist storms. One day he prepared a mixture of limestone and ordinary clay and took it to the bakery asking that it be baked in the oven at high temperature. When the result of his experiment came back, he noticed that the substance he had produced could harden in water and solidify to form a rock of the same kind as portlandite (from where the modern name of Portland cement). Using this substance, together with pozzolan coming from Italy, that for some reason he seemed to still trust unconditionally, he built a new lighthouse between 1756 and 1759, also known as the Smeaton tower. Had it not been for the underlying rocks, that were eroded by water, the lighthouse would still stand firmly where it was. It was actually removed in 1877.

Smeaton's studies were published right before his death and were probably read by James Parker, who filed a patent of dubious originality in 1796 and started the first business producing ``Roman cement'' -- this was its commercial name -- together with John Bazley White. Meanwhile in France, Louis Vicat was also studying hydraulicity: probably also influenced by Smeaton's work he published in 1818 the result of his \textit{Recherches expérimentales sur les chaux de contruction, les bétons et les mortiers ordinaires}, that was translated and read all over Europe. Also a civil engineer, he completed in 1824 the first bridge ever built with artificial cement, in Souillac, Southern France.

John Apsdin's patent, filed in 1824, and similar to Maurice Saint-Léger's, filed some years before with Vicat's contribution, marked the birth of another cement-producing company. Most importantly, it gave the opportunity to Apsdin's son, William, to apply his rebellious temperament to experimenting new techniques for making cement. One day he overcooked a sample, to the point that it had vitrified. Before throwing it away, he had the idea to pulverize what appeared as a block of burnt rock: he then noticed that, upon hydration, this magic powder formed a much stronger concrete than what he was used to. He had just invented clinker and modern Portland cement. Aspdin's recipe was then improved by his competitor Isaac Charles Johnson in the 1850's and, besides minor changes, is the one still in use in cement factories nowadays.

The fortune of cement in the following two centuries is there for all to see. Starting with the tunnel under river Thames, completed by Marc Brunel in 1843, cement has gained a leading position in infrastructure, public and residential building, and, with a bit more difficulty, in design and architecture. This was a long process that would not have been possible without the mechanization and centralization of the productive chain, initially motivated by entrepreneurs' necessity to get rid of the corporations of stone cutters. It was the 19th century and the second industrial revolution had just begun.

